\documentclass[9pt,twocolumn,twoside]{opticajnl}
\journal{opticajournal} 

\setboolean{shortarticle}{false}

\usepackage{lineno}
\usepackage{multirow}
\usepackage{svg}
\usepackage{placeins}
\usepackage{orcidlink}

\title{The Atacama Cosmology Telescope: Passband Measurements with an Analysis of Systematic Errors}

\author[1,*]{Thomas~Alford~\orcidlink{0000-0003-1942-1334}}
\author[2,3]{Rahul~Datta~\orcidlink{0000-0003-3853-8757}}
\author[4]{Erminia~Calabrese~\orcidlink{0000-0003-0837-0068}}
\author[5]{Steve~K.~Choi~\orcidlink{0000-0002-9113-7058}}
\author[6]{Mark~Devlin~\orcidlink{0000-0002-3169-9761}}
\author[7]{Shannon~M.~Duff~\orcidlink{0000-0002-9693-4478}}
\author[8]{Adriaan~J.~Duivenvoorden~\orcidlink{0000-0003-2856-2382}}
\author[9]{Rolando~Dünner~\orcidlink{0000-0003-3892-1860}}
\author[10]{Joseph~Golec~\orcidlink{0000-0002-4421-0267}}
\author[11]{Matthew~Hasselfield~\orcidlink{0000-0002-2408-9201}}
\author[12]{J.~Colin Hill~\orcidlink{0000-0002-9539-0835}}
\author[7]{Johannes~Hubmayr~\orcidlink{0000-0002-2781-9302}}
\author[1,2,3]{Jeffrey~McMahon}
\author[13,14]{Michael~D.~Niemack~\orcidlink{0000-0001-7125-3580}}
\author[15]{Lyman~A.~Page~\orcidlink{0000-0002-9828-3525}}
\author[15]{Suzanne~T.~Staggs~\orcidlink{0000-0002-7020-7301}}
\author[2]{Alex~Thomas~\orcidlink{0000-0001-9528-8147}}
\author[16]{Cristian~Vargas~\orcidlink{0000-0001-5327-1400}}
\author[13]{Yuhan~Wang~\orcidlink{0000-0002-8710-0914}}
\author[17]{Edward~J.~Wollack~\orcidlink{0000-0002-7567-4451}}

\affil[1]{Department of Physics, University of Chicago, Chicago, IL 60637, USA}
\affil[2]{Department of Astronomy and Astrophysics, University of Chicago, Chicago, IL 60637, USA}
\affil[3]{Kavli Institute for Cosmological Physics, University of Chicago, Chicago, IL 60637, USA}
\affil[4]{School of Physics and Astronomy, Cardiff University, The Parade, Cardiff, Wales CF24 3AA, UK}
\affil[5]{Center for Experimental Cosmology and Instrumentation, Department of Physics and Astronomy, University of California, Riverside, CA 92521, USA}
\affil[6]{Department of Physics and Astronomy, University of Pennsylvania, Philadelphia, PA 19104, USA}
\affil[7]{Quantum Sensors Division, NIST, 325 Broadway, Boulder, CO 80305, USA}
\affil[8]{Max-Planck-Institut für Astrophysik, Karl-Schwarzschild-Str. 1, 85748 Garching, Germany}
\affil[9]{Instituto de Astrof\'isica and Centro de Astro-Ingenier\'ia, Facultad de F\'isica, Pontificia Universidad Cat\'olica de Chile, Av. Vicuña Mackenna 4860, 7820436 Macul, Santiago, Chile}
\affil[10]{Department of Astronomy, University of Massachusetts Amherst, Amherst, MA 01003, USA}
\affil[11]{Center for Computational Astrophysics, Flatiron Institute, 162 5th Avenue, New York, NY 10010, USA}
\affil[12]{Department of Physics, Columbia University, New York, NY 10027, USA}
\affil[13]{Department of Physics, Cornell University, Ithaca, NY 14853, USA}
\affil[14]{Department of Astronomy, Cornell University, Ithaca, NY 14853, USA}
\affil[15]{Joseph Henry Laboratories of Physics, Jadwin Hall, Princeton University, Princeton, NJ 08544, USA}
\affil[16]{Mitchell Institute for Fundamental Physics and Astronomy, Texas A\&M University, College Station, TX 77843, USA}
\affil[17]{NASA Goddard Space Flight Center, 8800 Greenbelt Road, Greenbelt, MD 20771, USA}
\affil[*]{Corresponding author: tdalford@uchicago.edu}

\begin{abstract}
We present measurements of the spectral response of the four Advanced ACTPol (AdvACT) multichroic detector arrays, which observed in frequency bands centered near 30, 40, 90, 150, and 220 GHz. The passbands were measured with a Fourier transform spectrometer (FTS) combined with coupling optics that direct and match the output of the FTS to the AdvACT receiver. We perform optical simulations of the FTS and coupling optics in order to model frequency-dependent systematic effects in this system, examining the systematics of each individual measurement component in detail. We use these simulations to apply corrections to the measurements and evaluate the remaining systematic uncertainty. We combine these systematic estimates with statistical errors to determine the measured passbands and their uncertainties from our FTS measurement, which are typically $0.75$ to $1.5\%$ per detector array for the central frequency. We present the passband characteristics for all measurements made by ACT in a common framework. Additionally, we discuss in detail limits set by the systematics in our measurements as well as ways to minimize the systematics of each component to improve passband measurements in the future. \end{abstract}

\setboolean{displaycopyright}{false} 
\doi{} 
\dates{}
\begin{document}

\maketitle

\section{Introduction} \label{sec:introduction}

Measurements of the Cosmic Microwave Background (CMB) are key to our current understanding of cosmology and many astrophysical phenomena. Observatories are making rapid advances in sensitivity and frequency coverage, with a variety of bands at different frequencies. For example, Advanced ACTPol (AdvACT) \cite{Thornton_2016} observed in bands centered at 30, 40, 90, 150, and 220 GHz for a combined map depth of $10$ $\mu \mathrm{K}$ arcmin \cite{Naess_2025}, while Simons Observatory (SO) \cite{Simons} adds a band centered at 280 GHz to these frequencies and forecasts a combined map depth of $2.6$ $\mu \mathrm{K}$ arcmin \cite{Abitbol_2025} over an even larger sky area. Given the improved sensitivity of these surveys, we are at the point where the passband calibration can limit our ability to separate signals with different frequency dependencies and remove foregrounds. For instance, Simons Observatory has a number of science goals that crucially depend on precision characterization of the measured passband. Accurate characterization of the polarized foregrounds is essential for B-mode searches with the small aperture telescopes \cite{BandpassRequirements}. Likewise, for the large aperture telescope, proper bandpass characterization is critical for constraining foreground parameters \cite{LAT_bandpass_requirements_2024, Beringue_2025}, measuring lensing, tracing large-scale structure using the thermal Sunyaev-Zel’dovich effect, analyzing the CMB power spectrum, and placing limits on cosmic birefringence. Taking full advantage of all of these datasets is expected to require knowledge of the effective band centers at the 0.1\% level for some applications. The results presented in this work do not by themselves reach this threshold; however, they provide a necessary step toward it by identifying the dominant systematic effects in the measurements and quantifying their associated corrections and uncertainties.

\begin{figure}[t]
\includegraphics[width=.47\textwidth]{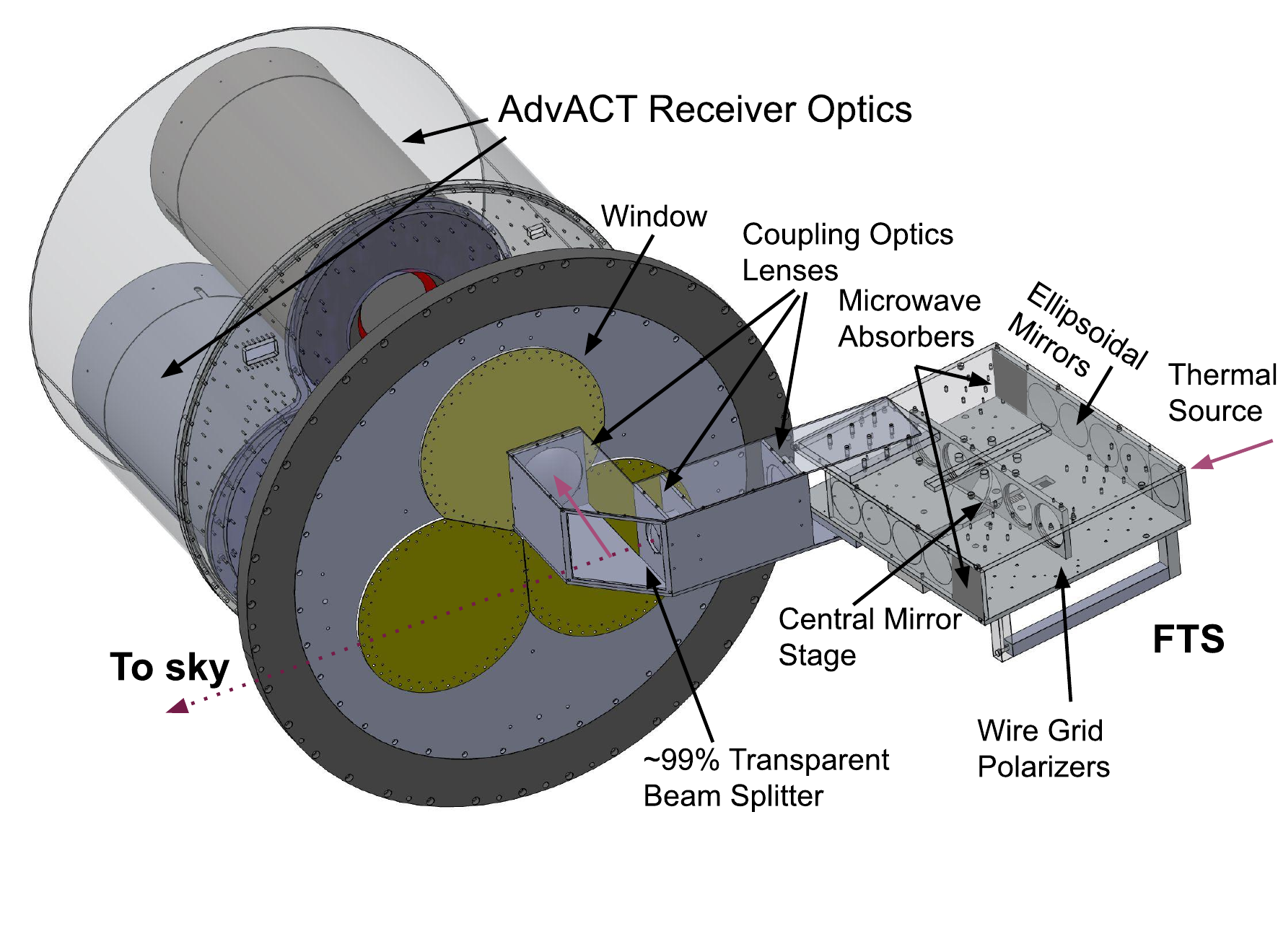}
\caption{CAD diagram of the FTS, coupling optics, and AdvACT optics tubes. The top of the FTS and coupling optics arm has been made transparent to show the internal components more easily. Most, $\gtrsim99\%$, of the power from the input thermal source is transmitted through the beam splitter to the sky to avoid detector saturation, while $\lesssim1\%$ is reflected toward the optics tubes. This view has been rotated to see features of the FTS and receiver more easily; the actual measurement was taken with the FTS pointing down towards the receiver at a 45 degree angle.} 
\label{fig:FTS_with_ACT_OT}
\end{figure}

The individual receivers in CMB polarimetry instrument systems typically have 30\% fractional bandwidth \cite{Thornton_2016, Simons, spt_spectra_2022}. With relatively wide bandwidth polarimetry channels, the calibration is especially challenging since relevant sources can have vastly different dependencies on frequency over the full spectral range of the instrument system. The flux of many foregrounds can be described by a power-law, $S_\nu \propto \nu^\beta$, with $\beta \sim -0.7$ for synchrotron emission and $\beta \sim 3.5$ for dust-like sources \cite{Page_2003, Choi_2020}. Precise knowledge of the observed amplitudes of these sources is essential for separating them out from maps of the CMB, which is generally done using internal linear combination procedures \cite{ILC_2003, MILC_2013} or parametric models that require precise knowledge of each sky component's frequency dependence \cite{Beringue_2025}. With knowledge of the anticipated source spectra, the amplitude for a beam-filling source can be computed by precisely understanding the effective frequency of observation or color correction. These both depend on integrals over the product of the passband and source spectra \cite{Page_2003}. Thus, achieving a 0.1\% knowledge of the band center requires careful calibration of the instrument response as a function of frequency. Systematic effects that shift the recovered frequency or produce frequency-dependent biases in the apparent amplitude response of the measurement must be carefully quantified and controlled.

\begin{table}[t]
\caption{FTS design parameters and specifications. }
\begin{tabular}{ll}
 \toprule
      Parameter & Value \\
 \midrule
    Mirror/Polarizing Wire Grid Center Spacing & 90 mm \\
    Mirror/Polarizing Wire Grid Diameters & 84.6 mm\\
    Dihedral Mirror Thickness & 0.1 mm \\
    Dihedral Mirror Maximum Center Offset & 75 mm \\
    FTS Width & 452.3 mm \\
    Beam Angle ($\theta$) & $11.3^{\circ}$  \\
    Étendue & $150$ $\text{mm}^2\text{-Sr}$ \\
    \bottomrule
   
\end{tabular}
\label{tab:FTS_specs}
\end{table}

The primary tool for bandpass calibration in CMB measurements is the Fourier transform spectrometer (FTS) \cite{optics}. Other instruments, such as vector network analyzers, have been used successfully to measure the frequency response of small numbers of coherent detectors \cite{Jarosik_2003}. However, the FTS is well suited to the calibration of large numbers of cryogenic detectors, especially when the end-to-end response of each detector including the feedhorn, orthomode transducer, and optical filter stack is desired. 

Fourier transform spectrometers create an optical delay between two input beams of light that are then recombined and passed through to a detector \cite{chamberlain1979principles}. The measured power as a function of optical delay (the interferogram) can then be Fourier transformed to recover the spectrum of the input source, or in our case the spectral response of a single-mode detector to a known input source.  A variety of FTS designs and measurement schemes have been used for CMB frequency characterization \cite{page_1994, matsuda_2019, karkare_2014, spt_spectra_2022, planck_spectra_2013}. For the Atacama Cosmology Telescope (ACT) we measured the passbands using an FTS with a Rayleigh-Jeans source input. The FTS was combined with coupling optics designed to match the beam of the FTS to the beam of the receiver optics. 

In this paper we present the system we used to calibrate the AdvACT receiver. We describe its design in Section \ref{sec:FTS_design} and we describe the modeling of systematic effects in Section \ref{sec:simulations}; the measurements of the AdvACT passbands are presented in Section \ref{sec:measurements}. We present the final passband results including statistical and systematic errors in Section \ref{sec:results}. We conclude with a discussion of the achieved performance and approaches that could be taken to improve the precision and accuracy in future calibration efforts.

\section{FTS and Coupling Optics Design} \label{sec:FTS_design}    
The FTS used in this work is based on the Primordial Inflation Explorer (PIXIE) design \cite{Kogut_2011} with additional coupling optics that match the outgoing beam from the FTS to the input of the AdvACT receiver. A flat beam splitter used in reflection limits the loading of the CMB detectors that would otherwise be saturated by the Rayleigh-Jeans source. A CAD model of the FTS and AdvACT receiver is shown in Figure \ref{fig:FTS_with_ACT_OT}.

\paragraph{FTS design}

Our version of the PIXIE FTS \cite{Pan_2019} contains four ellipsoidal mirrors on each side of the FTS, four central polarizing wire grids, and a movable central dihedral mirror. The input signal is a $\sim750$ K thermal emitter chopped at $\sim8$ Hz (labeled ``Thermal Source'' in Figure \ref{fig:FTS_with_ACT_OT}). Its radiation is divided by the wire grids into two orthogonally polarized beams that travel along the two sides of the FTS, reflecting off opposite sides of the dihedral mirror, before being recombined. Moving the dihedral mirror changes the optical path difference between the two beams. Their interference at the output produces the interferogram from which the spectral response is recovered.  

As in the PIXIE design, the FTS also has a second input port and a second output port. In our configuration the second input is a mirror covered with microwave absorber that emits as a 300 K blackbody. In general, the modulated part of the output interferogram is proportional to the difference between the two input signals. However, the 300 K input is not chopped, so its contribution is removed when the detector timestreams are demodulated at the chopper frequency. The second output port is terminated with microwave absorber and is not used.

\begin{figure}[t]
\includegraphics[width=\columnwidth]{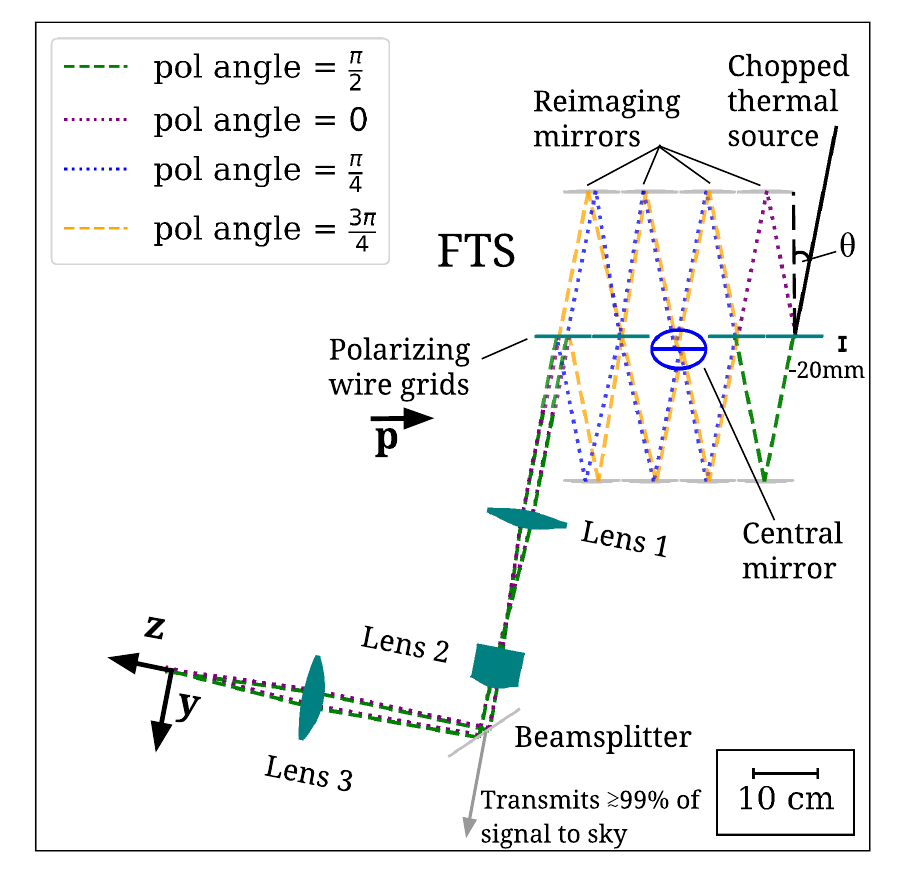}
\centering
\caption{An overhead view of the trace of a central ray through the FTS and coupling optics. The polarizing wire grids are oriented at angles of $\pi/2$, $\pi/4$, $\pi/4$, and $\pi/2$ radians with respect to the axis $\mathbf{p}$. Orthogonally polarized rays that travel through the same optical path are slightly offset from each other for clarity. The central mirror position is set to $h = -20$ mm.}
\label{fig:FTS_chief_trace}
\end{figure}

Each FTS mirror is a section of an ellipsoid whose foci lie at the centers of adjacent optical elements, re-imaging light from the previous mirror onto the next with a beam angle $\theta$, which is set by the dimensions of the FTS. The wire grid polarizers are aligned at angles $\pi/2$, $\pi/4$, $\pi/4$, and $\pi/2$ with respect to the FTS central axis (labeled $\textbf{p}$ in Figure \ref{fig:FTS_chief_trace}). The central dihedral mirror is mounted on a linear stage. A central mirror offset of $h$ generates an optical delay of $4 h \cos{\theta}$ between the two beams. The FTS component dimensions are detailed in Table \ref{tab:FTS_specs}. The FTS central mirror step size $\Delta h$ and total (two-sided) throw $H$ determine the Nyquist frequency 
 
\begin{equation} 
\nu_\textrm{max} = \frac{c}{8 \Delta h \cos{\theta}}
\label{eqn:FTS_nyquist}
\end{equation} 
and resolution of the spectra  
\begin{equation} 
\Delta \nu = \frac{c}{4 H \cos{\theta}},
\label{eqn:FTS_res}
\end{equation} 
where $c$ is the speed of light. The central dihedral mirror has a maximum displacement of $\pm 75$ mm, corresponding to a total throw of $150$ mm. During the measurements reported in this work, however, the mirror was scanned at most from $-35$ mm to $+35$ mm for a total throw of $70$ mm.

\begin{table}[t]
\caption{Design parameters of the HDPE lenses comprising the coupling optics.}
\begin{tabular}{lllll}
 \toprule
      Element & Thickness & R & $\kappa$ & $A_{4}$ \\
      & (cm) & (cm) &  & ($\textrm{cm}^{-3}$) \\
 \midrule
    Lens 1 surface 1$^a$ & 2.35   & 21.6 &   &     \\
    Lens 1 surface 2 & & $-11.4$ & $-1.857$ & $3\times10^{-5}$ \\
    Lens 2 surface 1$^a$ & 5.62   & $-7.1$ &  &   \\
    Lens 2 surface 2$^a$ & & $-5.2$ &  & \\
    Lens 3 surface 1 & 2.96   & 11.2 & $-0.960$ & $3\times10^{-5}$ \\
    Lens 3 surface 2 &    & $-27.8$  & $-6.539$ & $-2\times10^{-5}$ \\
    \bottomrule
   
\end{tabular}
\label{tab:coupling_optics}

\footnotesize{$^a$ Surface 1 of Lens 1 and both surfaces of Lens 2 are spheric surfaces. The rest of the surfaces are aspheric.}
\end{table}

\paragraph{Coupling optics}  

The coupling optics are a focal length reducer with an integrated beam splitter designed to transmit the full field of view of the FTS with high image quality. The coupling optics arm mounted to the FTS with output pointing directly at the AdvACT receiver is shown in Figure \ref{fig:FTS_with_ACT_OT}.  This system converts the FTS output f/5.6 to the f/1.6  input needed for the AdvACT receiver. This system is designed to illuminate clusters of about 13 feeds or 26 detectors per band at a time (out of $\sim 1000$ per focal plane array). Detectors outside the region illuminated by the FTS coupling optics remain open to the sky; however, some rays from these detectors may be vignetted by elements of the coupling optics or FTS mounting assembly. These detectors do not produce modulated interferograms and are therefore excluded from the analysis. A beam splitter reduces the signal by $\gtrsim 99\%$ to reduce the FTS signal incident on the receiver to hundreds of milli-Kelvin while allowing the receiver to receive the majority of its power from the sky.  This is critical as the detectors would saturate if the full FTS signal were used. This calibration approach implicitly assumes that inserting the FTS and coupling optics system does not significantly alter the optical termination seen by the focal plane, apart from introducing the intended modulated calibration signal.

These coupling optics are comprised of three high-density polyethylene (HDPE) lenses and a thin low-density polyethylene (LDPE) beam splitter. The lens surfaces include both spheric and aspheric surfaces defined as:
\begin{equation} 
    z = \frac{r^{2}/R}{1+\sqrt{1-(1+\kappa) r^{2}/R^{2}}} + A_{4}r^{4},
\end{equation}
where $z$ is the sag of the surface, $r$ is the radial polar coordinate, $R$ is the radius of curvature, $\kappa$ is the conic constant, and $A_{4}$ is a higher order term. Table~\ref{tab:coupling_optics} describes the shapes of the lenses.

\begin{figure}[t]
\includegraphics[width=.47\textwidth]{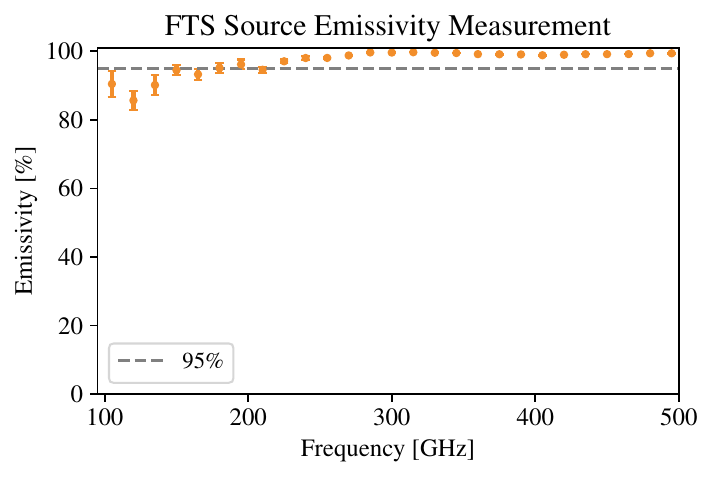}
\centering
\caption{Measurement of the ceramic heat lamp source emissivity at 300 K. This was determined by measuring the reflectance of a smooth portion of the lamp's surface. The measurements below 200 GHz likely underestimate the emissivity due to the difficulty of using our calibration source at these frequencies, but in the worst case the impact implied by the measurement is sub-percent for all bands.}
\label{fig:source_char}
\end{figure}

\paragraph{Chopped Thermal Source}

The thermal source for this FTS is a commercially available heat lamp behind a  chopper wheel. \footnote{The heat lamp can be found commercially at https://www.mcmaster.com with part number 3110K93.} 

The chopper modulates the signal at frequencies from a few to a few tens of Hertz, with 8 Hz being the default frequency. This was chosen to modulate above low frequency noise from fluctuations in the atmosphere. The heat lamp has a diameter of 120 mm, which fully fills the field of view. We operate at temperatures near $750$ K. Each FTS measurement was made with a step-and-integrate procedure by moving the central mirror over a set of discrete positions. The chopped signal was demodulated and integrated while the central mirror sat stationary at each position to build up the measured interferogram.

We tested this blackbody against a 77 K blackbody made using high-quality commercial microwave absorber submerged in a liquid nitrogen bath and found no indication of a departure from an RJ spectrum. We selected this heat lamp source for its ease of use. Additionally, the source was calibrated by measuring the reflectance as a function of frequency from its smooth surface and computing the emissivity via the Hagen-Rubens relation, $\epsilon = 1 - R$ \cite{BornWolf1999}. Figure \ref{fig:source_char} shows this measurement, which implies the emissivity to be greater than 80\% below 200 GHz and above 95\% for the higher frequencies. Within these two frequency regions the variation within a band with 30\% bandwidth is bounded by 5\% and 2.5\% linear slope and 10 GHz periodic oscillations at a similar level. We estimate the impact on the band centers and bandwidth to be sub-percent based on the model. A 1\% shift in the central frequency would require the emissivity to be below 40\% at 100 GHz assuming >95\% emissivity above 300 GHz and a similar falloff below. We note that the measurements below 200 GHz are likely an underestimate for the emissivity and an overestimate of the variation due to the difficulties of using the calibrator at these frequencies. For a future higher precision calibration an improved blackbody that is more accurately characterized will be required.

\section{Simulations of systematic effects} \label{sec:simulations}
\subsection{Simulation Procedure}

We use simulations to quantify effects that create systematic biases on the frequency or response of the system as a function of frequency. In this section we describe these simulations, while Section \ref{sec:measurements} will discuss how these effects were corrected and how the combined uncertainties were computed. These simulations are carried out using physical optics whereby ray-tracing is used to calculate the path length through our optical system along a given path and then multiple paths are summed to compute interference effects. The simulations are based on the code used in Liu et al. (2020) \cite{Liu:20} with a modification to accelerate the calculation of many frequencies by storing the path length and computing the integrals in a separate step \footnote{GitHub repository: https://github.com/tdalford/FTS-Simulation}. The simulations add additional traces through coupling optics between the FTS output and the AdvACT receiver.

\begin{figure}[t]
\centering
\includegraphics[width=0.47\textwidth]{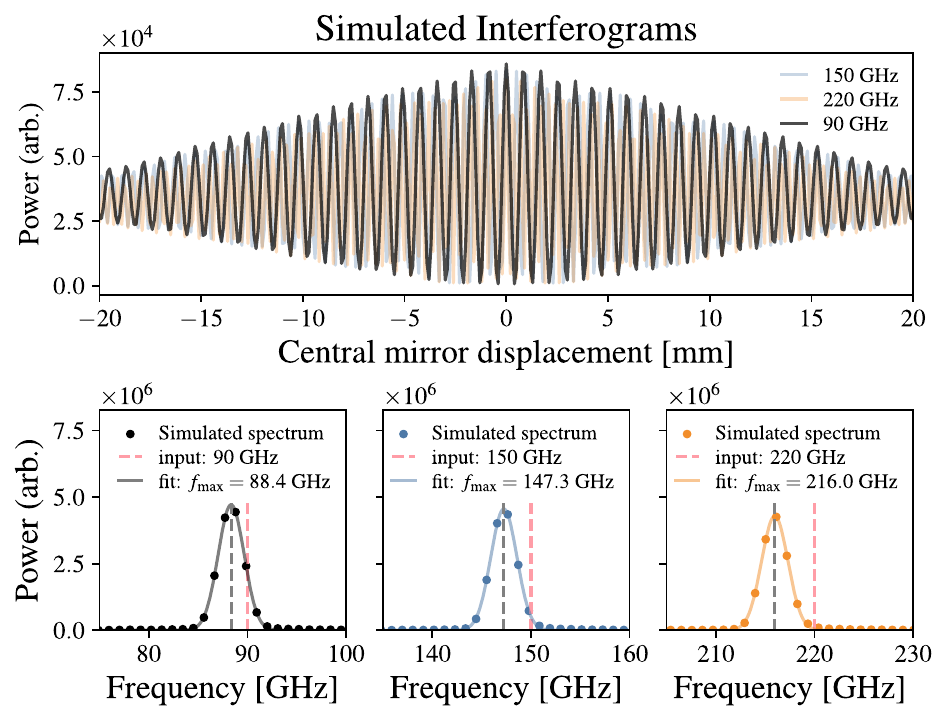}
\caption{\textit{Top:} Simulated interferograms at 90, 150, and 220 GHz for a centered detector and spectral line source. \textit{Bottom:} Individual spectra of the simulated interferograms, including a Gaussian fit to each. In each case the true frequency of the source is about a factor of $1.019 = 1 / \cos{\theta}$ times the fitted frequency, consistent with the optical delay relation $4 h \cos \theta$ for a central mirror displacement $h$. Small deviations from this factor are due to the extended angular distribution of the source. The recovered amplitude of each spectrum decreases slightly with increasing frequency due to decoherence effects from small path length variations in the FTS, representing the FTS transfer function. Because the detector shown here is centered, the recovered frequency shift is effectively negligible after accounting for the nominal $1/\cos\theta$ correction; for detectors displaced from the center, the recovered frequency shift is larger.}
\label{fig:simulated_interferograms}
\end{figure}

\begin{figure}[t]
\centering
\includegraphics[width=.47\textwidth]{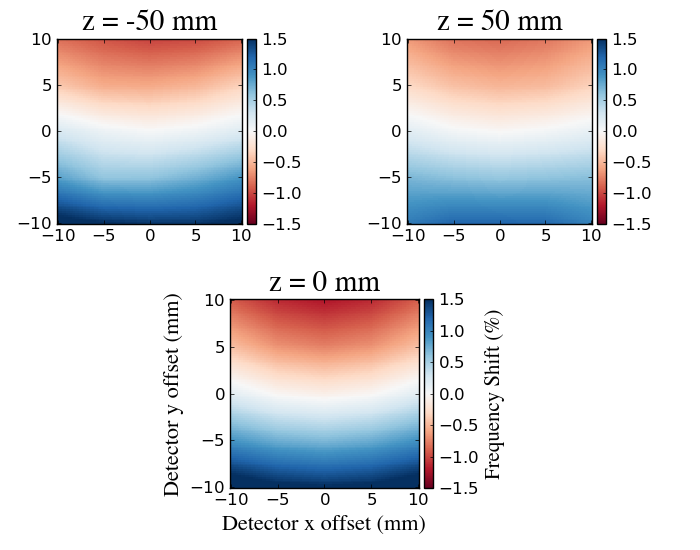}
\caption{Simulated fractional frequency shifts across the focal plane at perfect (z = 0) and maximum (z = $\pm 50$ mm) alignments of the FTS to the receiver. The detector $x$ and $y$ offsets are relative to the center of the FTS pointing on the focal plane. The maximum $x$ and $y$ offsets of $\pm 10$ mm correspond to the vignetting limit of the FTS and match the amount seen in measurements. Plotted is the frequency shift at a frequency of 150 GHz. In practice simulations over all possible misalignments and starting frequencies between 15 and 300 GHz are computed, although the fractional frequency shift is nearly independent of starting frequency. The frequency shifts have primarily a dependence on the $y$ position of the detector on the focal plane relative to the FTS pointing center but do vary slightly with $x$.}
\label{fig:freq_shift_examples}
\end{figure}

\begin{figure}[t]
\centering
\includegraphics[width=.47\textwidth]{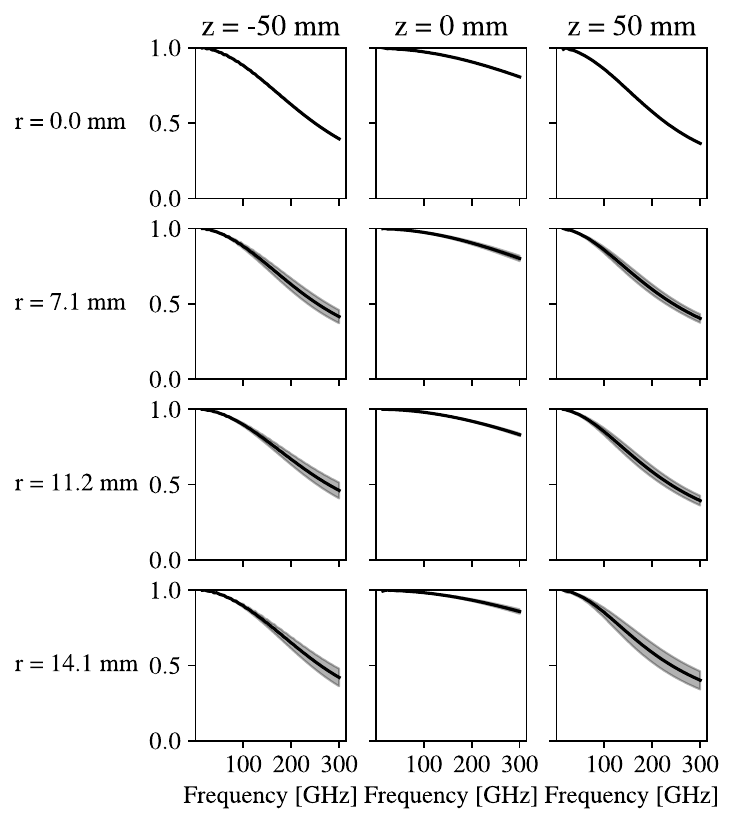}
\caption{FTS transfer functions between 15 and 300 GHz simulated and normalized for perfect FTS z alignment ($z = 0$, middle column) along with maximum z misalignments ($z = \pm 50$ mm, left and right columns) in either direction. The simulated corrections are plotted as a function of radius from the coupling optics focus. The radially symmetric corrections have errorbars plotted as shaded regions that are subdominant to the correction dependence on $r$ and $z$.} 
\label{fig:defocus_color_correction}
\end{figure}

\begin{figure*}[t]
\centering
\includegraphics[width=\textwidth] {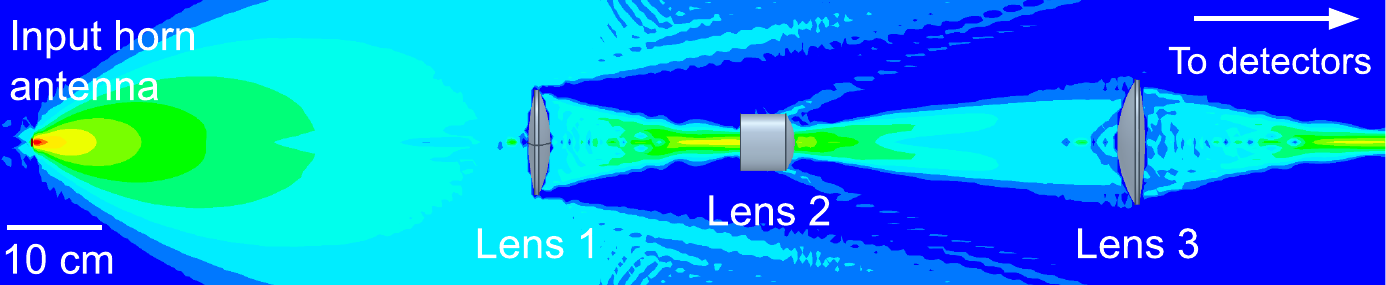}
\caption{Simulation of the coupling optics lens transmission using Ansys High-Frequency Structure Simulator (HFSS) software. The simulation includes a centered source transmitting at 75 GHz that is modeled as a pyramidal horn antenna pattern that propagates through to the focus of the coupling optics. To find the transmitted power through the lenses, we perform simulations at a range of source positions matched to the full extent of the thermal source and at a range of frequencies up to 100 GHz. Then we sum up the total power which resides within a detector feed radius from all source positions at each frequency.} 
\label{fig:coupling_optics_HFSS_sim}
\end{figure*}

These simulations launch rays starting at the thermal source as the input to the FTS and are treated as a single linearly polarized state. They are propagated through the optical system, storing the optical path length traveled and losses (due to reflection from or transmission through a polarizing wire grid) as they encounter each element, until finally reaching a position on the detector focal plane. Any rays that miss an optical element are discarded. At this point the rays would be combined in the feedhorn of a single detector. To simulate this, for a given feed position, all rays that land outside the outer feedhorn radius are not included in the computation. The detector field is then formed by summing the complex contributions of rays that couple to the detector feed, with each ray's wave amplitude weighted by the beam pattern of the detector feedhorn. A $5 \times 5$ square grid of possible detector feed positions is analyzed to probe the dependence of the spectral response on the detector position. The analyzed feed positions extend to a maximum displacement of $10$ mm, corresponding to the vignetting limit of the FTS. A trace of the central ray with an offset central mirror position $h = -20$ mm is shown in Figure \ref{fig:FTS_chief_trace}.

This calculation is repeated for a number of dihedral mirror positions to simulate an interferogram at individual frequencies, which is then Fourier transformed to calculate the spectrum for a monochromatic input.  We fit a Gaussian to each monochromatic spectrum. The parameters in the Gaussian fit are interpreted as the amplitude and recovered central frequency for this input. This procedure is repeated over a range of frequencies. These Gaussian fit parameters are then used to construct the transfer function, or response as a function of frequency, and frequency shifts of the FTS.  These are studied as a function of position and with respect to alignment errors. An example of spectra generation and parameter estimation from simulated interferograms is shown in Figure \ref{fig:simulated_interferograms}.

To simulate the extended source we choose starting points evenly spaced over the 120 mm circular thermal source, with a total of 625 starting positions used. At each point the rays are treated as being emitted with an isotropic directional dependence. We use $n = 4500$ starting rays at each point. We calculate spectra for frequencies evenly spaced between 15 and 300 GHz. Since the rays are incoherent in phase when emitted from different starting positions, for a given detector position we sum the interferograms together from each starting position before transforming to the final spectrum. 

\paragraph{Coordinate System}
We define a coordinate system to quantify the impact of relative errors in the optical alignment of the FTS and the receiver on passband measurements. The $z$ axis is defined by the direction the central ray travels as it exits the coupling optics, i.e. normal to the center of last lens and focal plane. The $y$ direction lies in the coupling optics / FTS plane and is orthogonal to the optical axis $z$. The $x$ direction is then perpendicular to $z$ and $y$. These are illustrated in Figure \ref{fig:FTS_chief_trace}.

During the measurements presented below, the alignment of the coupling optics relative to the receiver was not precisely controlled.  While the x-y alignment is naturally encapsulated within the field of view of the FTS, the z offset represents an uncontrolled variable that will factor into systematic uncertainties.   We estimate that the alignment along this axis was only controlled to $\pm 50$ mm and varied between measurements. To evaluate this effect we simulated 30 possible z-distance offsets of the FTS between these values, $z = +50$ mm and $z = -50$ mm (we denote $z = 0$ as perfect alignment).

\subsection{Systematic Effects}

We use the simulations described above and additional calculations to model the impact of four classes of systematic effects: 

\begin{enumerate} 
\item[1.] Frequency shifts arising from variations in the path through the FTS; 
\item[2.] Color corrections due to optical effects in the FTS;
\item[3.] Color corrections due to the transmission of the lenses;
\item[4.] Color corrections due to the reflectance from our thin film beam-splitter.
\end{enumerate}

An alternative way to find the total frequency shift and color correction for this setup would be to measure a calibration receiver with a relatively flat or well-known response. This work instead uses simulations to model each individual component’s correction along with its uncertainties. This approach allows for more  precise knowledge of the contributions from each component of the optical system and helps identify areas for improvement in future FTS measurements.

\subsubsection{FTS Frequency Shifts} \label{subsubsec:freq_shift}

The output of the FTS depends on the interference of light.  Any physical realization of an FTS includes optics that accommodate multiple paths through each arm. This leads to geometric effects that limit the output resolution and induce biases which amount to frequency shifts in the recovered spectrum.  

We use the simulations described above to calculate this frequency shift as a function of detector position (x-y) and z-alignment. The frequency shift was simulated for a range of starting frequencies between 15 and 300 GHz but the fractional shift was found to have negligible variation with starting frequency. 

The results of these simulations are presented in Figure \ref{fig:freq_shift_examples}. These contour plots summarize the spatial dependence of the fractional frequency shifts as a function of $x$, $y$, and $z$ offsets in the alignment of the FTS relative to the receiver focus.  These indicate a weak dependence on the focal alignment along the boresight direction ($z$) leading to a bias of at most 0.2\% due to our $\pm 50$ mm alignment accuracy.  Larger errors are associated with in-plane displacements (the $x$-$y$ plane). Here we find that the largest shifts occur along the $y$ direction.  These roughly linear variations in the frequency shift arise because detector offsets from the FTS focus in $y$ cause the angles of incidence through the FTS to vary.  This detector displacement relative to the FTS focus leads to $\sim 1.5\%$ shifts in the band center.  This implies a several percent variation in band-center when illuminating multiple detectors. Measurements when multiple detectors were illuminated are consistent with this prediction. It is the dominant systematic effect in the determination of band-centers with this FTS.

\subsubsection{FTS Transfer Function} \label{subsubsec:fts_defocus}
The FTS transfer function is the ratio of the output power relative to the power presented at its input port. Generally, this function is frequency dependent due to variations in the path length delays within the spectrum of rays defining the beam, diffraction from the edges of optical elements, decoherence arising from diffusive surface scattering, and other processes. In our simulations we examine the first process. The variations in optical delay between rays in different parts of the beam lead to a frequency dependent reduction in the amplitude.  Shorter wavelengths will yield stronger decoherence effects as the wavelength approaches the size of these small optical delay fluctuations. This effect thus causes a measured passband to attenuate more at higher frequencies. The transfer function was simulated by combining the amplitudes of the fits to each individual frequency spectrum as shown in Figure \ref{fig:simulated_interferograms}. This effect is dependent on the detector position in the focal plane and the z alignment of the FTS relative to the detectors. This contribution to the transfer function is roughly symmetric with radius from the coupling optics focus. Figure \ref{fig:defocus_color_correction} shows the simulated transfer functions for various radii at z = 0, +50, and -50 mm alignment. Near the focus this effect amounts to 5\% changes over an octave band.  However, this can grow to up to 50\% per octave when significantly out of focus.

\begin{figure}[t]
\centering
\includegraphics[width=.47\textwidth]{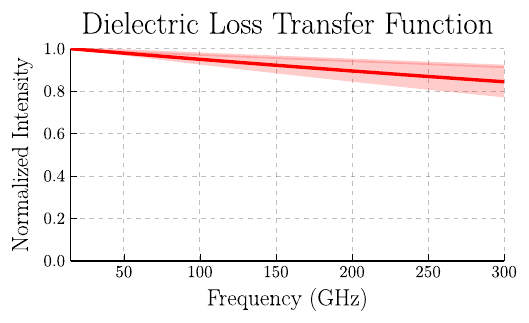}
\caption{Transfer function due to dielectric losses in the HDPE lenses of the coupling optics. Uncertainties are plotted as the shaded region and are calculated assuming a loss tangent that can vary from  $1 \times 10^{-4}$ to $3 \times 10^{-4}$.}
\label{fig:loss_tangent}
\end{figure}

\begin{figure}[t]
\centering
\includegraphics[width=.47\textwidth] {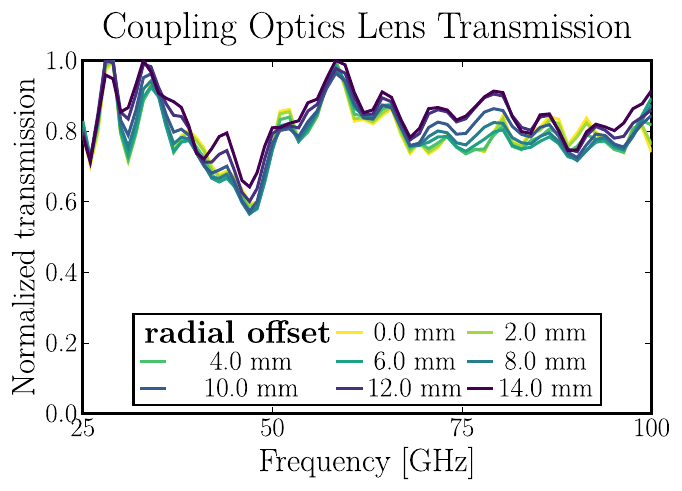}
\caption{Transmitted power through the coupling optics to a detector feed as a function of frequency and from different radii from the center of the FTS pointing, as computed from Ansys HFSS simulations. A quadratic fit has been removed from each transfer function to account for the frequency-dependent beam size of the input pyramidal horn antenna (this spuriously modifies the amount of power reaching the first lens and propagating through the simulation). The simulation is used to estimate the relative magnitude of spectral interference effects in this refractive coupling optics system. Although there are small differences between the pyramidal horn antenna pattern used in the simulation and the actual antenna illumination pattern of the ACT feeds, these differences are not expected to significantly affect the relative interference effects found.} 
\label{fig:coupling_optics_transmission}
\end{figure}

\begin{figure}[t]
\centering
\includegraphics[width=.45\textwidth]{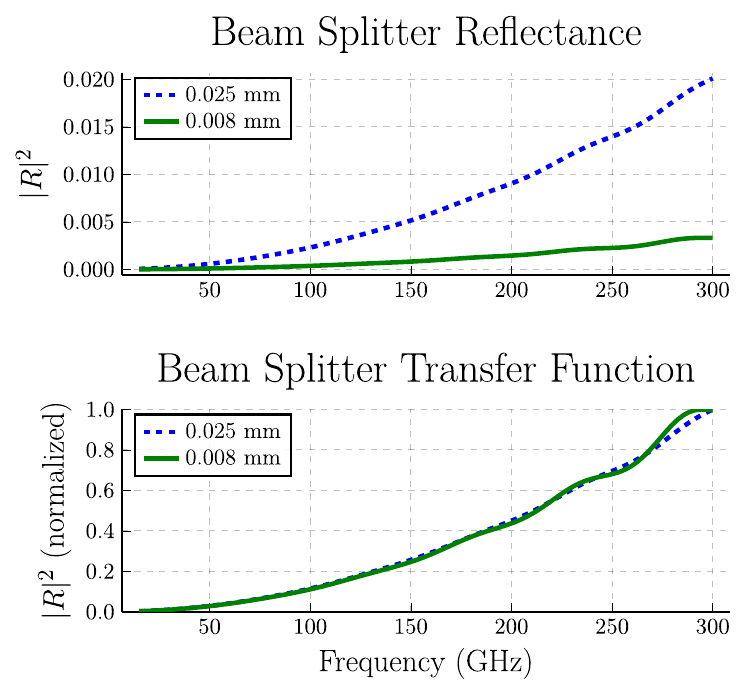}
\caption{Simulated beam splitter reflectance (top) and spectral shape correction (bottom) functions for thicknesses of 0.008 mm and 0.025 mm between 15 and 300 GHz. The transfer functions are effectively identical between the two cases, while the absolute reflectances differ by roughly a factor of 6.}
\label{fig:thin_membrane}
\end{figure}

\subsubsection{Coupling Optics Transfer Function }\label{subsubsec:lens_transmission}

The three HDPE lenses used in the coupling optics, which do not have any anti-reflection coating, absorb light and create multi-path interference due to reflections from each of their surfaces leading to an additional transfer function.  

The intensity transmission coefficient for light traveling through the lenses is approximated by (ignoring surface reflections) 
\begin{equation}
\label{eqn:coupling}
T = e^{-\tan \delta \frac{2 \pi n}{\lambda} l},
\end{equation}
where $n$ is the index of refraction of the lenses ($n = 1.52$), $\tan \delta$ is the loss tangent of the material (HDPE) that is estimated to be between $1 \times 10^{-4}$ and $3 \times 10^{-4}$ \cite{HDPE_loss_tangent, HDPE_loss_proceedings}, and $l$ is the distance traveled through the lenses. 

While the total path length $l$ through the lenses varies, we estimate the transmission using the mean path length of 96 mm.  Fixing this value has a negligible effect on the estimated transmission since variations in the dielectric losses from path length deviations are smaller than those arising from the assumed uncertainty in the loss tangent. This transfer function and its uncertainty are shown in Figure \ref{fig:loss_tangent}. The amplitude of this loss is 7\% per octave.   

Interference effects in refractive systems are a more significant concern as they introduce a variety of coherent features in the resulting spectra. To estimate the amplitude of these effects, we simulate the transmission of the coupling optics system in Ansys Electronics High-Frequency Structure Simulator (HFSS) \footnote{http://www.ansys.com/Products/Electronics/ANSYS-HFSS}. The simulation does not include the polarizing wire grids or mirrors from the FTS, as incorporating these components would increase the computational requirements beyond our current computing capabilities. We instead assume that the dominant interference effects arise from the refractive coupling optics lenses and limit our simulation to these components. The HFSS model also excludes the beam splitter, whose effect is treated separately in the next section.

Figure \ref{fig:coupling_optics_HFSS_sim} shows the HFSS model with a color scale proportional to the field strength. The simulation uses a pyramidal horn on the left as the input source of light propagating through the coupling optics system before coming to a focus on the right.  The real part of the $z$ component (the propagation direction of the simulation) of the Poynting vector is summed over a radius equivalent to the radius of a detector feed and weighted by a Gaussian approximation to the beam of the feed. We combined simulations from 25 - 50 GHz and 40 - 100 GHz which use two different waveguide dimensions for the input pyramidal horn. We could not simulate higher frequencies because the computation time scales with the cube of the frequency. However we expect effects at higher frequencies to be qualitatively similar to the effects seen in this simulation. Although the pyramidal horn does not exactly reproduce the illumination pattern of the detector array, the simulated transmission is used only to estimate relative spectral interference effects. Differences in feed taper and spillover are therefore not expected to significantly affect the final calibration.

A significant amount of the power propagating from the pyramidal horn in the simulations misses the first lens. Since the Gaussian beam response from this input horn narrows with increasing frequency this causes a spurious frequency-dependent trend in the power received at the detector feed. We account for this by fitting and removing a quadratic term from the transmission function.

Figure \ref{fig:coupling_optics_transmission} shows the results of this simulation. This transfer function includes oscillations at a scale of approximately 5 GHz and amplitude of typically 10\% for this frequency separation. The phase and amplitude of these variations depend on the coordinates $r$ and $z$ of the detector with respect to the optics position. Correcting for this transmission exactly would require precise knowledge of the distributions of $r$ and the distance $z$ for the measured detectors. Averaged over a passband, the oscillations produce a small mean shift in band center compared with their variation over $r$ and $z$. For most bands we therefore do not correct for interference. Instead, we use the simulations to set an uncertainty on the band centers (Section \ref{subsubsec:apply_corrections}) and to interpret features in the measured spectra. The exception is PA7 f030, whose narrow band makes the mean shift dominant, so we correct for it explicitly. The dielectric loss component of the transfer function (Figure \ref{fig:loss_tangent}) is corrected for in all bands.

\subsubsection{Beam Splitter Transfer Function}\label{subsubsec:thin_membrane}

The $\gtrsim 99\%$ transparent beam splitters used to prevent saturation have frequency-dependent transmission. LDPE of thickness 0.008 mm was used as the beam splitter for the 30/40 GHz array while 0.025 mm thickness was used for the higher frequencies. We simulated the reflection coefficient from 15 to 300 GHz for LDPE of thickness 0.008 mm and 0.025 mm using HFSS. Figure \ref{fig:thin_membrane} shows the results of the simulations. Going from 0.025 mm to 0.008 mm membrane thickness yields a factor of six less reflectance; however, when normalizing these reflection functions, they have the same frequency dependence up to a shift in the overall amplitude. This increase in reflectance with increasing frequency causes the measured passband to have increased amplitude at higher frequencies. The uncertainty in this correction is negligible since we are not sensitive to shifts in the overall amplitude.

\begin{table}[t]
\centering
\setlength\tabcolsep{2pt}
\caption{Details of individual measurements of the Advanced ACTPol detector arrays. The FTS throw indicates the total distance the central mirror travels to both sides of the FTS. Each "run" corresponds to a pointing position of the FTS which illuminates up to $\sim$ 26 detectors in the focal plane, where many runs are needed to properly sample a large portion of the detector array. The total number of spectra indicates the number of detectors with physical spectra measured with the FTS over all observing dates and runs.}
\begin{tabular}{|l|c|c|c|c|c|}
\hline
Array / & Measurement & FTS Step & FTS & Total \# & \# \\
 Frequency & Dates (mm/yy) & Size & Throw & Spectra & Runs \\
\hline
  PA4-f150 & 12/16, 12/17, & \multirow{2}{*}{0.1 mm} & \multirow{2}{*}{40 mm} & 179 & 26 \\
  PA4-f220 & 12/19  &  &  & 193 & 24 \\
  \hline
  PA5-f090 & \multirow{2}{*}{12/17, 12/19}  & \multirow{2}{*}{0.15 mm} & \multirow{2}{*}{70 mm} & 53 & 16 \\
  PA5-f150 &  &  & & 150 & 16 \\
  \hline
  PA6-f090 & \multirow{2}{*}{12/17, 12/19}  & \multirow{2}{*}{0.15 mm} & \multirow{2}{*}{70 mm} & 83 & 26 \\
  PA6-f150 &   &  & & 248 & 27 \\
  \hline
  PA7-f030 & \multirow{2}{*}{01/21} & \multirow{2}{*}{0.375 mm} & \multirow{2}{*}{70 mm} & 29 & 15 \\
  PA7-f040 &   & &  & 48 & 19 \\
\hline
\end{tabular}
\label{tab:measurement_attrs}
\end{table}

\begin{figure*}[t]
\centering
\includegraphics[width=\textwidth]{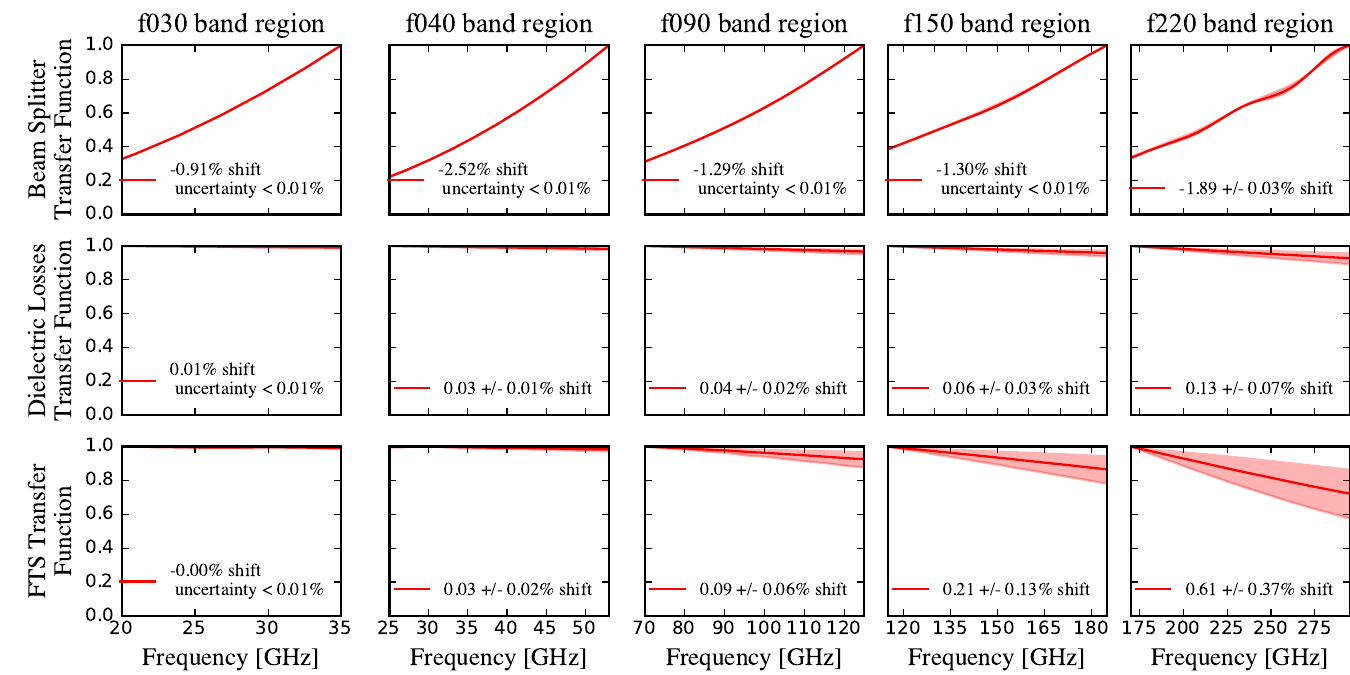}
\caption{Comparison of systematic transfer function (Section \ref{subsubsec:thin_membrane}, \ref{subsubsec:lens_transmission}, and \ref{subsubsec:fts_defocus} from top to bottom) magnitudes and uncertainties over each band region. Each transfer function plotted is normalized to a maximum of unity within an interval of frequencies chosen to simulate frequency intervals for the actual passbands. The shaded regions are minimum and maximum bounds of the transfer function distributions over the given interval. To calculate the percent shifts in band centers we sample the transfer function distribution and apply the corresponding correction, given by the inverse of the transfer function, to an original measured average passband at each frequency region. The shifts reported are the mean and standard deviation of each sample. Any shift or uncertainty calculated to be smaller than 0.01\% is reported as <0.01\%.
The percent shifts for the band centers in actuality depend on the choice of band used but these numbers are intended to be rough estimates given real band shapes.}
\label{fig:color_correction_compars}
\end{figure*}

\section{AdvACT Passband Measurements} \label{sec:measurements}

From 2016-2021, summer measurement campaigns were carried out to measure the AdvACT passbands (measurements in the preceding years characterized earlier ACTPol passbands \cite{Datta_2016}; we focus on the AdvACT passbands in this work). These campaigns measured the four separate AdvACT cameras called polarization arrays (PA for short). Each of these cameras was multichroic and covered a set of neighboring frequency windows. These were PA4 (AdvACT 150/220 GHz array), PA5 and PA6 (AdvACT 90/150 array), and PA7 (AdvACT 30/40 GHz array) \cite{AdvACT_henderson_dets_2016, Koopman_2018, Thornton_2016}. The measurement date, number of spectra for each array, FTS measurement parameters, and number of observing runs are summarized in Table \ref{tab:measurement_attrs}. These data permit tests of the spatial and temporal variation in the passbands. Each array was measured with the FTS moved to roughly 15-30 different positions across the array.

\subsection{Data Selection}

The FTS and coupling optics only illuminate up to 26 detectors per band at a given pointing.  The first step in data analysis is to identify these channels and reject the channels not directly illuminated by the FTS. The set of illuminated detectors was identified by computing the spectra for each channel and then constructing a signal-to-noise metric from the mean in the band of interest divided by the RMS from frequencies well separated from this band.  

Detectors for which this statistic was above five were retained for further consideration.   Detectors that were spatially separated from the cluster illuminated by the FTS were also rejected.

\subsection{Data Processing}

The data that passed the above cuts were then analyzed to produce cleaned spectra. First, a polynomial fit was removed from the timeseries data to remove any large-scale drift in the interferograms. This polynomial was chosen to be 5th-order, which corrects any non-physical drifts in the interferograms without biasing the results (the bias in band center due to this removal is below $0.1\%$). 
A small number of glitches from jumps in the detector time streams were found in some channels. Most channels had no jumps but about 10\% of our data had a few such glitches. These glitches were masked and the missing values were replaced with linear interpolation between neighboring points. Next, the interferogram was apodized about the zero path difference point and the data was Fourier transformed to obtain the spectrum. The mean FTS calibration factor from the simulations ($\frac{1}{.981} \approx 1.019$), matching the expected geometrical factor from the optical delay formula $4 h \cos{\theta}$, was then multiplied to the frequencies.

\begin{figure}[!hbt]
\centering
\includegraphics[width=.47\textwidth]{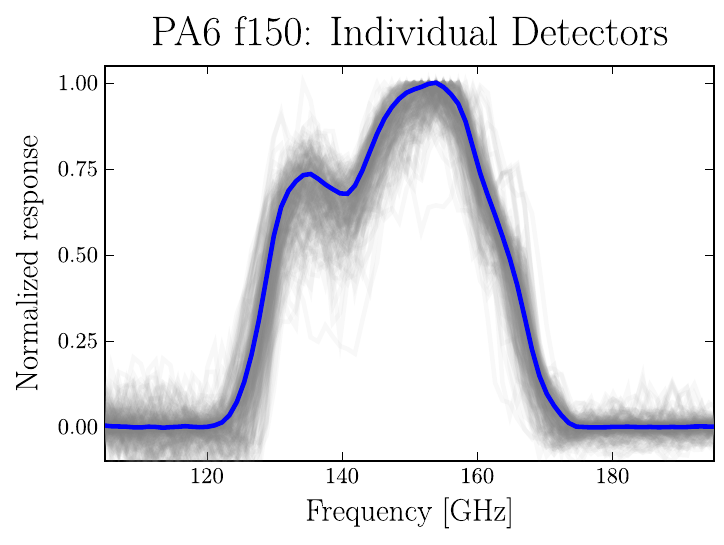}
\caption{All 248 measured passbands of the PA6 150 GHz detectors are individually plotted in gray. The final noise-weighted average is over-plotted in blue.}
\label{fig:all_dets_PA6}
\end{figure}

The passbands were tested both with and without applying the Mertz phase correction procedure \cite{mertz_corrections} to the data, which includes fitting and removing a linear gradient from the phase frequency dependence to account for a measured off-center interferogram. The results were nearly identical aside from increased noise and more detectors cut in the data with the correction applied. This occurred because the phase fits were unstable for low signal-to-noise detectors and for data containing glitches that were not fully removed. As a result, we chose not to apply the Mertz phase correction, which leaves a positive noise bias in the data. This, if unaccounted for, would bias the band center or width calculations, especially for measurements with low SNR. We correct for this by subtracting a constant baseline from each spectrum. The baseline term is determined by averaging the spectrum over a region far away from the main passband region and any associated harmonics.

The passband was then normalized to have a maximum value of 1, and the noise was estimated by computing the standard deviation of the data outside the passband region. This noise estimate was used in the signal-to-noise (SNR) weighting when combining multiple channels. Simulations were used to confirm that the bias introduced by these processing steps is less than 10\% of the uncertainties we present below.  

We recomputed the signal-to-noise of these processed spectra and rejected any bands with SNR < 15. We then averaged all the passband measurements for each band weighted by their SNR squared. In some cases, due to overlapping FTS positions, a single detector had multiple passband measurements between observing runs for the same array. In these cases we took the highest SNR measurement as the passband measurement to use.  The spectrum for each detector array is reported as the noise-weighted mean of these data and the statistical error is reported as the standard error on this mean within each frequency bin. For example, Figure \ref{fig:all_dets_PA6} shows individual processed spectra combined with the final noise-weighted average for the PA6 150 GHz array.

\begin{figure}[t]
\centering
\includegraphics[width=.44\textwidth]{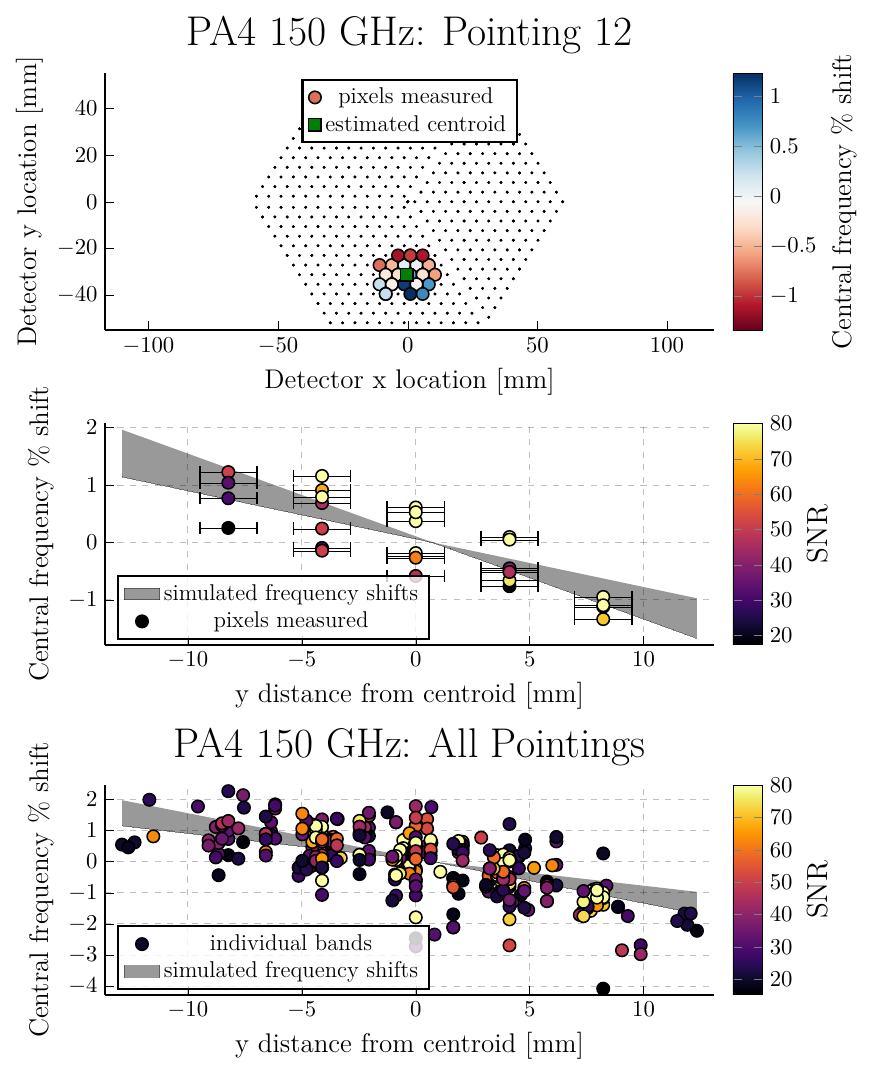}
\caption{An example of the FTS frequency shift seen in actual passband data.
\textit{Top:} Pixel positions with measured passbands from run 12 (out of 26 total runs) of the PA4 150 GHz measurement. The color scale represents the central frequency shift of each pixel measured relative to the average passband. The green square shows the centroid, or center of the cluster of measured detectors, for this run.
\textit{Middle:} Run 12 PA4 150 GHz frequency shifts which have been plotted as a function of the y distance from the centroid, along with the expected shifts from simulations. The errorbar for each point is the estimated standard error on the centroid y position. The color scale shows the SNR of each measurement.
\textit{Bottom:} Similar pixel data but now including passbands from all PA4 150 GHz pointings combined. The trend in the measurements is consistent with expected slope from the frequency shift simulations in Section \ref{subsubsec:freq_shift}. The horizontal errorbars have been removed for visibility.}
\label{fig:freq_shift_cluster}
\end{figure}

\subsection{Systematic Bias Corrections and Error Propagation}

\subsubsection{Obtaining Final Systematic Corrections}
\label{subsubsec:apply_corrections}
There are four corrections that need to be applied to the final passbands to take into account the systematic effects discussed in Section \ref{sec:simulations}: the FTS frequency shift correction (\ref{subsubsec:freq_shift}), the FTS transfer function (\ref{subsubsec:fts_defocus}), the coupling optics lens transmission correction (\ref{subsubsec:lens_transmission}), and the beam splitter correction (\ref{subsubsec:thin_membrane}). 

The mean coupling optics transmission from dielectric losses and beam splitter transfer function lead to multiplicative biases that are straightforward to correct and for which we can estimate systematic uncertainties.  

The FTS transfer function correction and FTS frequency shift correction are more complex since these are dependent on the detector position relative to the FTS. The FTS was moved to 15-30 different positions across each detector array. However, the exact central pointing position of the FTS onto the focal plane was not known during measurement. It can only be estimated roughly from the centroid of the pixels measured for each pointing. This estimate is limited by the low number of active pixels in each pointing and exacerbated by the fact that some detectors were not active during a given measurement. We therefore carry conservative systematic uncertainties in both of these corrections. 

We evaluate the FTS transfer function over all radii and $z$ alignment positions and make a correction based on the mean of these functions. Because the $z$ offsets during the measurements are unknown and possibly correlated between runs, we conservatively estimate the variance by sampling either the largest (at $z=+50$ mm) or smallest (at $z=0$ mm) correction with 50\% probability. Figure \ref{fig:color_correction_compars} shows the beam splitter, coupling optics dielectric loss, and FTS transfer function corrections in each individual band region, along with the resulting uncertainties on the central frequencies.

\begin{table}[t]
\renewcommand{\arraystretch}{1.3} 
\centering
\caption{Uncertainty composition of the AdvACT passband measurements, with each column reporting the contribution to the total error on the central frequency of the passband. Column 2 shows the statistical error on the noise-weighted average passband. Column 3 shows the uncertainty estimate of all multiplicative corrections, including contributions from \ref{subsubsec:fts_defocus} and \ref{subsubsec:lens_transmission} (\ref{subsubsec:thin_membrane} has negligible associated uncertainty). Column 4 shows the residual RMS of all passband central frequencies after correcting for the y dependence of the frequency shift \ref{subsubsec:freq_shift}. Column 5 shows the upper limit of the uncertainty expected from the full coupling optics transmission in \ref{subsubsec:lens_transmission}.}

\begin{tabular}{ccccc}
\toprule
Array /&Statistical&Transfer&Residual&Coupling\\
Band&Error& Function& RMS& Optics\\
\midrule
PA4-f150 & 0.1\% & 0.1\% & 0.7\% & 0.3\%\\ 
PA4-f220 & 0.1\% & 0.3\% & 1.5\% & 0.3\%\\ 
PA5-f090 & 0.1\% & 0.1\% & 0.9\% & 0.3\% \\ 
PA5-f150 & 0.1\% & 0.1\% & 0.8\% & 0.3\%\\ 
PA6-f090 & 0.2\% & 0.1\% & 1.1\% & 0.3\%\\ 
PA6-f150 & 0.1\% & 0.1\% & 0.6\% & 0.3\%\\ 
PA7-f030 & 0.2\% & 0.01\% & 0.8\% & 0.8\%\\ 
PA7-f040 & 0.5\% & 0.02\% & 3.3\% & 0.7\%\\ 

\bottomrule
\end{tabular}
\label{tab:band_error_budget}
\end{table}

\begin{figure*}[t]
\centering
\includegraphics[width=\textwidth]{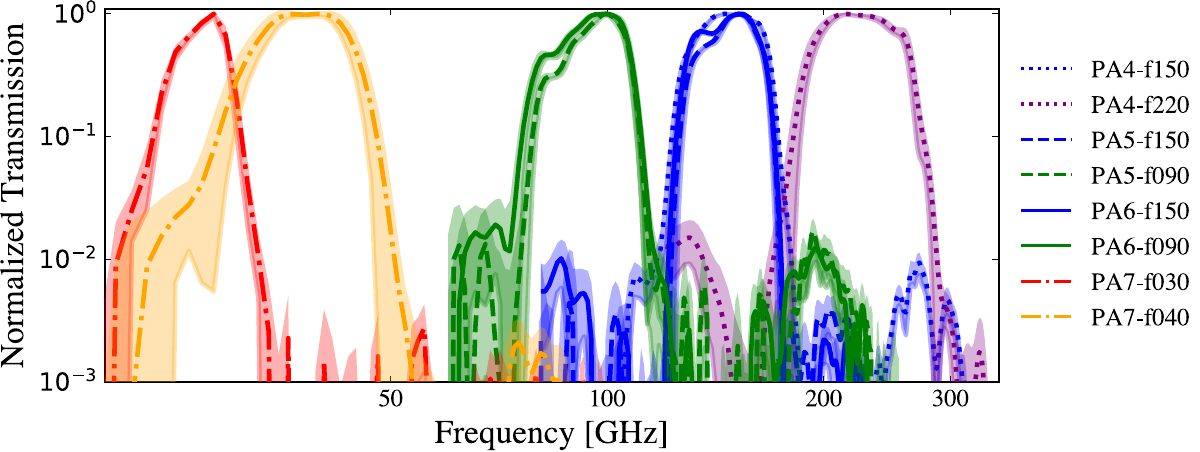}
\caption{AdvACT PA4 (150 and 220 GHz), PA5 (90 and 150 GHz), PA6 (90 and 150 GHz), and PA7 (30 and 40 GHz) passbands. The shaded regions show combined statistical and systematic (spectral shape and frequency shift corrections) errors for each band. Low frequency noise swiftly increases due to the beam splitter correction and is not included in the bands plotted. The measurement floor of approximately 1\% is expected to be due to white noise in the FTS measurements and not real. The harmonics at roughly twice the band frequencies of the 90 and 150 GHz bands are believed to be spurious features as a result of nonlinearities and aliasing. The resolution of the spectra as computed from Equation \ref{eqn:FTS_res} is 1.9 GHz for PA4 and 1.1 GHz for the other arrays. The measurements have not been smoothed otherwise.} 
\label{fig:passbands}
\end{figure*}

The largest systematic uncertainty on band center determination is the FTS frequency shift, which can shift the bands by up to $1.5\%$ in either direction.  We measure this effect in our data by computing the band centers of individual detectors and plotting that against the y distance from the center of the detector centroid in a given pointing.   Figure \ref{fig:freq_shift_cluster} shows this for the PA4 150 GHz data.  A clear slope is visible at a level consistent with the expected variation.  We correct this effect by removing the best-fit slope in the y-direction.  The dominant error in applying this correction is uncertainty in the true detector positions.  This leads to an intrinsic variation in the bands between detectors and at different pointings. We do not separately account for the systematic uncertainty on this correction. We report the uncertainty on this correction as the residual RMS of the bands post correction. This accounts for any intrinsic errors in pointing estimation. Another source of uncertainty arises because we did not sample the full detector array. This combined with any potential systematic shifts in the passbands across the wafer would yield a biased wafer-averaged estimate. However, using the residual RMS as the final uncertainty also accounts for this possible variation assuming enough distinct sampling regions. 

Finally, the frequency dependence of interference effects in the coupling optics must be accounted for. As mentioned in \ref{subsubsec:lens_transmission}, we use the simulations to place limits on the effects of interference in the lenses and add a conservative uncertainty determined by the variance in the band shifts seen when applying each simulated transmission function in $r$ and $z$. In general, the variance of the in-band transmission functions dominates the mean correction for each band, with the exception of PA7 f030. For all bands besides PA7 f030, the variance is treated as an RMS scatter to the frequencies and is added in quadrature to the residual RMS to the bands from the previous paragraph. Since the simulation only goes up to 100 GHz, the 150 and 220 GHz array passbands use the 90 GHz band uncertainty ($0.3\%$) as the evaluated uncertainty. However, PA7 f030 is explicitly corrected for and adds an additional variance equal to the amount shifted from the mean correction in quadrature. 

Table \ref{tab:band_error_budget} compares the sources of uncertainty on the passband central frequency for each AdvACT detector array. The dominant remaining source of uncertainty is the residual RMS among the detectors, although we note that applying the best-fit slope correction to each array reduces the RMS by up to 25\%.

\subsubsection{Propagation of the Errors}
\label{subsubsec:error_prop}

We multiply each of the mean transfer functions together to arrive at the net correction which is divided out of the passband measurement. Our main types of uncertainties are systematic error bounds on these transfer functions, statistical errors of the measured average, and RMS frequency shifts. We combine all of these errors together by Monte Carlo sampling from the distributions of each until the final error distribution converges. 

In each Monte Carlo iteration we start with the final measured average passband. We first sample a statistical error of the average and apply it to the band. We then apply realizations of each transfer function. The FTS transfer function is chosen with equal probability from the maximum ($z=+50$ mm) or minimum ($z=0$ mm) simulated correction (Section~\ref{subsubsec:apply_corrections}). The dielectric loss transfer function is computed with a loss tangent chosen with equal probability from the bounds of its estimated range, $1\times10^{-4}$ or $3\times10^{-4}$. The beam splitter correction is applied as a fixed correction and is not sampled, since its uncertainty is negligible (Section \ref{subsubsec:thin_membrane}). Finally, we shift the band by a frequency offset drawn from a Gaussian distribution whose width is the quadrature sum of the residual RMS and the coupling optics RMS. We compute the central frequency and bandwidth of each resulting passband, eventually building up a distribution used to estimate uncertainties. We compute uncertainties on both the central frequencies and bandwidths in this manner, although we note that the frequency shift uncertainty has no effect on the bandwidth.


\begin{figure*}[ht]
  \centering
  \includegraphics[width=\textwidth]{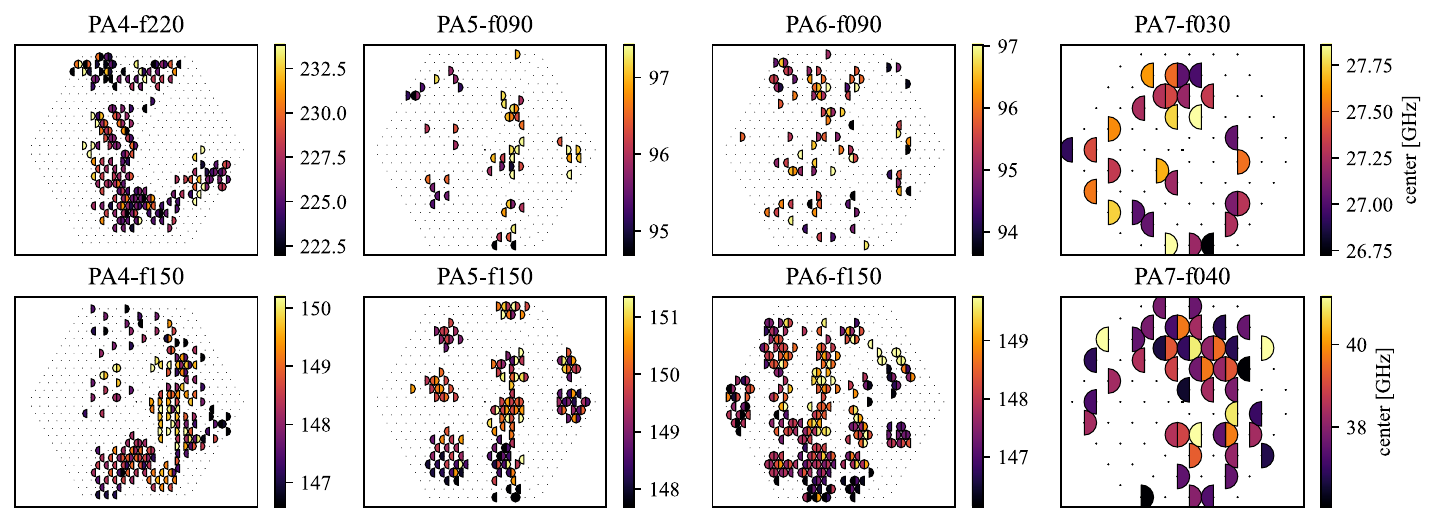}
  \caption{Spatial variation of the passband central frequencies for each detector array. Orthogonally polarized detectors with the same passband residing at the same pixel are mapped to polarization groups `A' and `B'. `A' detectors are plotted with a left half-circle while `B' detectors are plotted with a right half-circle. The typical fraction of functional detectors per array was greater than 70\%, with reduced spatial coverage in the plots from detectors missed or otherwise cut in the measurements. There is a slight radial trend seen primarily in the 150 GHz data where passbands near the edge of the array have $\sim 2\%$ lower central frequency than passbands at the center. The maximum and minimum values for each colorbar are bounded by $95\%$ intervals in its corresponding dataset to avoid outlier effects.} 
\label{fig:spatial_variation}
\end{figure*}

\begin{table}[ht]
\centering
\caption{Final Band Attributes and Propagated Uncertainties}
\begin{tabular}{llll}
\toprule
Array &      Band &      Center [GHz] &    Bandwidth [GHz] \\
\midrule
  PA4 &   150 GHz &   148.5 $\pm$ 1.1 &   35.8 $\pm$ 1.4 \\
  PA4 &   220 GHz &   226.1 $\pm$ 3.5 &   64.9 $\pm$ 2.5 \\
  PA5 &    90 GHz &   96.4 $\pm$ 0.9 &   18.9 $\pm$ 1.0 \\
  PA5 &   150 GHz &   149.3 $\pm$ 1.3 &   28.1 $\pm$ 1.4 \\
  PA6 &    90 GHz &    95.3 $\pm$ 1.1 &   23.1 $\pm$ 1.0 \\
  PA6 &   150 GHz &   147.8 $\pm$ 0.9 &   31.2 $\pm$ 1.4 \\
  PA7 &    30 GHz &    27.4 $\pm$ 0.3 &    4.3 $\pm$ 0.4 \\
  PA7 &    40 GHz &    38.3 $\pm$ 1.3 &   12.4 $\pm$ 0.8 \\
\bottomrule
\end{tabular}
\label{tab:band_centers}
\end{table}

\section{Results} \label{sec:results}

A plot of the final AdvACT passbands and their fully propagated errors is shown in Figure \ref{fig:passbands}. The measurements have statistical noise floors approximately at the 1\% level due to white noise in the FTS measurements, with a gentle frequency dependence due to the decreasing beam splitter reflectance at lower frequencies. Low-frequency out-of-band data is low signal-to-noise and is not shown. We have no evidence for spurious couplings at low frequencies, a conclusion that is further supported by thick grill filter measurements that bound the integrated couplings to below 1\% in the low frequency out-of-band region.

The passbands measured with this setup determine the frequency response of the AdvACT detectors to a beam-filling source in the Rayleigh-Jeans (RJ) limit. The power measured from an RJ source is weighted only by the detector's frequency bandpass response with no other frequency dependence. The figures and results shown in the main section of this work use the central frequency defined as 

\begin{equation}
     \nu_0 = \frac{\int \nu \tau(\nu) \,d \nu}{\int \tau(\nu) \,d \nu},
\label{eq:RJ_center_freq}     
\end{equation}
where $\tau(\nu)$ is the passband response at a given frequency $\nu$. 

We characterize the bandwidth using the signal bandwidth. This computation is independent of the source spectrum and is defined as (see Appendix \ref{appen:formalism})

\begin{equation}
     \Delta \nu = \int \tau(\nu) \,d \nu,
\end{equation}

where we set $\tau(\nu)$ to be normalized to have a maximum of one. 

Table \ref{tab:band_centers} shows the final band centers and widths, along with their uncertainties, for an RJ source spectrum. However, in practice, the effective central frequency of the passband depends on the spectrum of the celestial source observed. Appendix \ref{appen:formalism} introduces the formalism for computing full effective frequencies and color corrections for different types of sources. Complete results tables for different ACT arrays, including effective frequencies for multiple types of celestial and point sources, as well as the various factors used in the results, are included in Appendix \ref{appen:results}. We note that while the effective frequency can vary significantly depending on the celestial source, its uncertainty varies very little ($\lesssim 0.1$ GHz).

\begin{figure}[ht]
\centering
  \includegraphics[width=.47\textwidth]{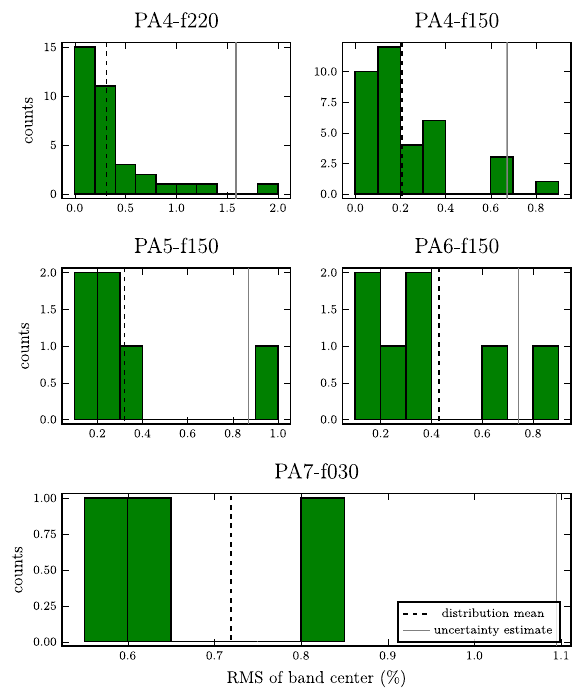}
  \caption{Temporal variation in the passband centers for each array. We found all detectors with at least three measurements and computed the noise-weighted standard deviation of the measured central frequencies. Plotted is the percent uncertainty for each such detector. The PA5 90 GHz, PA6 90 GHz, and PA7 40 GHz arrays had no such detectors with at least 3 different measurements. Plotted as vertical lines are the mean values and the percent uncertainty estimates of the band centers for each array. In all cases the scatter is smaller than the estimated uncertainty; there is no evidence for temporal variation in the band centers.} 
\label{fig:temporal_variation}
\end{figure}

\subsection{Integration Bounds and Aliased Harmonics}

When computing these band centers and bandwidths, a choice must be made about which frequency range to integrate over. Ideally, the band could be measured perfectly and the central frequency and bandwidth computed as an integral over all frequencies. However, spurious trends and noise artifacts in our data would otherwise bias the results. For instance, there are several above-band features that mimic harmonics of the 90 and 150 GHz bands and are aliased by our step-and-integrate data acquisition strategy. These are attributed to nonlinearity in our detectors for the relatively large signals from the FTS. Lab tests using similar detectors for Simons Observatory showed similar spectral features--- measurements using a coherent source confirmed these to be systematics associated with compression in our bolometric detectors \cite{carlos_latrt_paper, Harrington_2020}. We thus do not include these harmonics in our computations of band centers and bandwidth.


An additional spurious noise source in the data arises from the beam splitter correction, which substantially increases the noise at low frequencies. To avoid biasing our results, we exclude most of the below-band low-frequency region from the integration; this is also justified because that region lies below the expected waveguide cutoff of each band. Thus, to minimize the effects of these low and high frequency biases we set integration bounds within at most 10 GHz of the band edges.

\paragraph{Possibility of Blue Leaks}


Constraints are placed on out-of-band leakage from the amplitudes of point sources seen in the maps. Dust-like point sources have steeply rising spectra ($\beta \sim 3.5$) and are therefore particularly sensitive to blue leaks. A comparison of ACT PA4 f220 and South Pole Telescope (SPT) 217 GHz dusty point-source amplitudes \cite{Reichardt2021} limits any shift in the dust-like band center from high-frequency leakage to less than $3\sigma$ of the band-center uncertainty reported here. This assumes that SPT does not have a common systematic with ACT. This is a reasonable assumption since the two instruments use somewhat different optical filter stacks. This constraint can be used to place limits on a variety of leakage models. For example, this rules out a broad $1\%$ leak in integrated power of the main band near its second harmonic between 400 and 500 GHz. Other modes for leaks can also be bounded with this method. However, smaller leaks, or leaks at lower frequencies, cannot be definitively ruled out. Such leakage could still bias the effective frequency by amounts larger than the uncertainties reported in this work. Constraining these leaks will be particularly important for future passband measurements requiring an even smaller error margin. The results and uncertainties presented in this work account for systematic errors from the FTS itself and do not include possible out-of-band leakage.

\subsection{Spatial Variation of the Bands}

The band centers as a function of detector position are shown in Figure \ref{fig:spatial_variation}.  This figure also shows which detectors were measured in each array.  The typical fraction of detectors measured ranges between 10\% and 50\%. In some cases, a lack of spatial coverage is a result of missing detectors in the arrays themselves-- though the fraction of functional detectors was typically greater than 70\%. Orthogonally polarized detectors that reside at the same pixel and have the same passband region are separated into two polarization groups `A' and `B'. To differentiate between these, the figure shows two detectors per pixel with the half-circle on the left `A' detectors and on the right `B' detectors. We separately tested consistency between A and B detectors and found no significant difference in their passbands up to measurement uncertainty. 

In most cases any possible trends in the spatial variation of the band centers are obscured due to intrinsic measurement uncertainty across the detector array. However, after applying the systematic corrections as detailed in \ref{subsubsec:apply_corrections}, the data exhibits a small radial dependence, primarily in the 150 GHz data. Detectors near the center of the array have roughly 2\% higher central frequency than detectors at the far edge. Such a radial trend is possible due to differences in the deposited dielectric thickness over a wafer and has been observed in similar types of arrays \cite{carlos_latrt_paper} but is not yet well understood. 

\subsection{Temporal Variation of the Bands}

In some cases, array measurements were repeated multiple times with the FTS in the same position. In other cases, the extended beam size of the FTS output resulted in the same detector being measured from multiple different FTS positions. These repeated measurements span both daily timescales within a single measurement campaign and yearly timescales between campaigns. We compare all such measurements to set limits on the time variability of the passbands. Because the number of repeated measurements is limited, we do not separately characterize variability on different timescales.

For detectors measured at least three separate times, we compute the noise-weighted standard deviation of the central frequencies to obtain an estimate of the variability. The results are shown in Figure \ref{fig:temporal_variation}. In the PA5 90 GHz, PA6 90 GHz, and PA7 40 GHz arrays we found no such detectors with at least three measurements.  For the remaining detectors we find variations in the band centers to be 1\% or less in almost all cases. Since our measured uncertainties for the bands are all at a similar or larger scale than this variation we interpret this as a non-detection consistent with our known systematic uncertainties. 

\subsection{Use in ACT DR6 Analyses}
The passband results produced in this work have been incorporated into multiple ACT DR6 analyses. In the component-separated maps of \cite{Coulton_2024}, the frequency dependence of each astrophysical component is integrated over the measured passbands to compute its expected response in each frequency channel. The passband uncertainties are also propagated to evaluate the impact of instrumental systematics on the resulting maps. In the power-spectrum likelihood analysis of \cite{Louis_2025} and the constraints on extended cosmological models in \cite{Calabrese_2025}, the measured passbands are similarly used to determine the band-integrated foreground responses. The measurement uncertainties set the prior widths for additional passband-shift parameters, which are sampled jointly with cosmological and foreground parameters for the final analysis. We note that the values reported in Table \ref{tab:band_centers} differ by 0.1 - 0.2~GHz per band from the bandpass-shift priors used in Table 4 of \cite{Louis_2025} due to minor updates to the analysis in this work. The posterior sampled shifts (Figure 26 of \cite{Louis_2025}) are all within the $1\sigma$ band uncertainties from this work with the exception of PA4-f220, which yielded a 1.4 $\sigma$ shift upwards relative to the prior. 

\section{Conclusion} \label{sec:conclusion}

We have presented measurements of the AdvACT passbands using a system comprised of a PIXIE-style FTS and coupling optics.  We have quantified the statistical and systematic errors.  We find that the band-center uncertainties due to the FTS measurement are at most 1.5\% for all bands except PA7 40 GHz, with the statistical error largely subdominant to the systematics.

Future experiments may require passband calibration accuracies approaching 0.1\%. This will be particularly challenging for astrophysical sources with substantially different spectral energy distributions. For high resolution CMB analyses, it will also be essential to account for the frequency dependence of the beam across the instrumental passband and its uncertainties \cite{Giardiello_2025}.

Achieving this level of accuracy will require improved control of both systematic and statistical uncertainties. The major systematic limitations on this measurement come from (1) the refractive lenses in the coupling optics, (2) alignment of the FTS, including precise knowledge of the FTS and detector pointing positions, and (3) possible out-of-band leakage. Addressing these systematics and reducing the statistical uncertainty represent tractable challenges.

The coupling optics systematics can most directly be mitigated by using a system of reflective coupling optics designed to create a similarly focused output beam with neutral frequency dependence. We are working on designing such reflective optics for future large aperture and small aperture telescope calibration.

Future measurements will also require precision alignment and determination of the FTS distance to the cryostat, particularly to reduce the frequency shift uncertainty. This can be performed either by aligning the FTS once by maximizing the input signal to the detectors and then fixing the axial distance in subsequent measurements, or by probing the effect and uncertainty of the axial distance through multiple measurements at a range of possible $z$. For instance, future measurements for the SO large aperture telescope plan on using a gantry system to precisely control and measure $z$ while taking measurements. The off-axis radial distance uncertainty can be reduced through precision detector beam-maps with the FTS alongside spectra measurements, which would allow precise determination of $r$ for each working detector. The FTS frequency shift as a primary source of uncertainty can also be addressed by closing the FTS beam down to a much smaller size by placing a small aperture over the source, making sure that we only measure detectors very close to the coupling optics focus at the cost of reduced illumination.

While the statistical errors of the main band region are close to the $0.1\%$ level in some cases, they can be further improved by sampling the full detector wafer and taking multiple samples per detector. This has the advantage of further probing the spatial dependence of the wafer passbands while sampling the entire wafer for an unbiased estimate of the average. Additionally, multiple samples of individual detectors will allow us to probe passbands at the single detector level. This may be required for precision cosmology given the possible spatial dependence of the passband over the detector wafer.

Finally, in order to constrain shifts in the effective band center from out-of-band leakage, the noise level of the measurement will have to be orders of magnitude deeper than the current 1\% noise floor, requiring much longer integration time with the FTS and a complete understanding of the baseline removal. To assist with these strict requirements, other calibration methods such as the frequency-tunable laser source (FLS) \cite{FLS_paper} are being developed to provide alternative calibration techniques to the FTS. The FLS performs exceptionally well out of band by being able to constrain leakage at the level of $0.01\%$ or less, while the FTS currently provides better measurements of the main band shape. Combining these techniques will allow us to perform stringent tests of the band systematics and reach even lower error thresholds.


\appendix

\section{Effective Frequencies and Color Corrections}
\label{appen:formalism}

The power that lands on a detector in a single polarization is given by 
\begin{equation}
P=\frac{1}{2}\iint I_\nu(\hat{n}) \eta(\nu) \tau(\nu ) A_e(\hat{n}, \nu) d\nu\,d\Omega_\nu,
\label{eq:power}
\end{equation}
where $\nu$ is the frequency, $I_\nu$ is the spectral or surface brightness with units of ${\rm W/m^2\,Hz\,sr}$, and $\tau(\nu)$ is the passband normalized to unit peak.
We take $A_e(\hat{n}, \nu)$, where $\hat{n}$ is associated with $d\Omega_\nu$,
to be the effective area assuming a lossless antenna. The optical coupling to the detectors ensures that only a single spatial mode of radiation is measured and thus $\int d\Omega_\nu\, A_e(\hat{n}, \nu) = \lambda^2$ at each frequency.  We account for the combined loss in all elements of the system with the efficiency $\eta(\nu)$ and in the following assume it is independent 
of frequency.\footnote{We generally adopt the notation of \cite{condon_ransom_2016} and \cite{planck_IX}.} When the source is a beam-filling Planck function,
$I_\nu=B_\nu(T)$, then in the Rayleigh-Jeans (RJ) limit\footnote{In the Rayleigh-Jeans limit, $h\nu<<k_BT$ where $h$ is Planck's constant and
$k_B$ is Boltzmann's constant.}
\begin{eqnarray}
P&=&k_BT \eta \iint \frac{\nu^2}{c^2} \tau(\nu )A_e(\hat{n}, \nu)  d\nu\,d\Omega_\nu \nonumber\\
&=& k_BT \eta \int \tau(\nu ) d\nu =
k_BT \eta \Delta\nu. 
\label{eq:one_mode_power}
\end{eqnarray}
In the last expression, $\Delta\nu=\int \tau(\nu) d\nu$ is the signal bandwidth of the receiver. It is always less than or equal to the Dicke bandwidth \cite{dicke_1946}, 

\begin{equation}
\Delta_{\scriptscriptstyle D}\nu =  \frac{ \bigg[  \int_{0}^{\infty} \tau(\nu)d\nu \bigg]^2}{ \int_{0}^{\infty} \tau^2(\nu)d\nu }
= \frac{ (\Delta_{\scriptscriptstyle}\nu)^2}{ \int_{0}^{\infty} \tau^2(\nu)d\nu },
\end{equation}
that enters into computing the sensitivity of a radiometer.

Note that in equation~\ref{eq:one_mode_power} the power depends only on $T$, $\eta$, and $\Delta\nu$ and does not depend on the frequency of the measurement. 
When the passband is measured with a Fourier Transform Spectrometer (FTS), the source is almost always in the RJ limit. Thus the measurement gives the contribution from each sub-band, $\delta\nu$,
of $\Delta\nu$ weighted only by $\tau(\nu )$ and no other powers of frequency. In this derivation we will assume that $\tau(\nu )$ is measured with a FTS in one spatial mode in the RJ limit of a source at constant temperature.

\subsection{Celestial Sources}
\label{sec:cel_sources}
For our purposes, the primary distinction between different sources is whether they are beam-filling or point sources; we leave a full $\ell$ dependent formalism outside the scope of this work. For beam-filling sources, the particulars of the beam forward gain should drop out whereas for point sources, the forward gain is an important part of the characterization. 

The ACT CMB temperature maps are calibrated in terms of deviations from a Planck emitter at $T_{CMB}$ with surface brightness 
\begin{eqnarray}
B_\nu(T_{CMB})&=& \frac{2h\nu^3}{c^2}\frac{1}{\exp{(h\nu/k_BT_{CMB})}-1} \\
&=& b_0T_{CMB}\frac{x^3}{e^x-1}, 
\label{eq:rj2}
\end{eqnarray}
where $x=h\nu/k_BT_{CMB}$ and $b_0= 2k_B^3T_{CMB}^2/(hc)^2=9.905\times10^{-19}$J/Km$^2$. The RJ limit is 
$b_0T_{CMB}x^2$.

Maps of diffuse sources are given in terms of surface brightness $I_\nu$. We are only sensitive to variations in $I_\nu$  which, in terms of a CMB calibrated map, are given by
\begin{equation}
\delta I_\nu = \delta T_{CMB} \frac{\partial B_\nu(T)}{\partial T}\bigg |_{T_{CMB}} \equiv b^\prime_\nu\delta T_{CMB} 
\label{eq:dcmb}
\end{equation}
so that
\begin{equation}
b^\prime_\nu =  
\frac{\partial B_\nu(T)}{\partial T}\bigg |_{T_{CMB}}=
 b_0\frac{x^4e^x}{(e^x-1)^2}.
\label{eq:bprime}
\end{equation}
In the RJ limit, $b^\prime_{\nu} = b_0x^2 = 2k_B\nu^2/c^2$. 


We often want to convert between fluctuations in brightness temperature $T_{RJ}$ and $T_{CMB}$. From Equation~\ref{eq:rj2} and its RJ limit we find
\begin{equation}
t_\nu^\prime\equiv\frac{\delta T_{RJ}}{\delta T_{CMB}} =  \frac{x^2e^x}{(e^x-1)^2}.
\label{eq:tprime}
\end{equation}
In the RJ limit, $t_\nu^\prime = 1$.\footnote{In \cite{Choi_2020},
$g_2(\nu)^{-1}= b_\nu^\prime$,  $g_1(\nu)^{-1}=t_\nu^\prime$, and $f(\nu)= \tau(\nu)$.}

The ACT temperature maps, for example, currently do not have information below a multipole of $\ell=200$, corresponding to angular scales larger than $0.5^\circ$. Thus any measure of a diffuse source is a variation from a local average as indicated in Equation~\ref{eq:dcmb}.
In the following, though, we use the notation $I_\nu$ to denote a diffuse source at frequency $\nu$ and implicitly recognize it as a fluctuation over some unspecified spatial scale.

In general, diffuse sources can be characterized as
\begin{equation}
I_\nu=I_{\nu_r}(\nu/\nu_r)^\beta
\label{eq:dsrcdef}
\end{equation}
where for synchrotron emission $\beta=-0.7$, for free-free emission $\beta=-0.1$, for RJ emission $\beta=2$, and for  dust-like galactic emission $\beta\approx3.5$. The $\nu_r$ is a convenient reference or pivot frequency.
More specific versions of this generic parameterization include the CMB, diffuse dust, the cosmic infrared background (CIB), and the Sunyaev-Zel'dovich effects \cite{SZ_1972}.

Point source emission is also generally characterized with a power law as $S_\nu=S_{\nu_r}(\nu/\nu_r)^\beta$ where $S_\nu$ is the flux in Janskys. The conventions for $\beta$ are the same as for beam-filling sources but the terminology is slightly different. For example, ``flat spectrum" radio sources have $\beta\approx 0$. Because of the variety of processes that produce these power laws, there is a large variation in $\beta$ especially at millimeter wavelengths where the  
indices span from roughly $\beta= -0.7$ to $\beta= 3.5$; thus classification by spectral index can be misleading. In general, $\beta$ can vary over a band but we do not treat this case.

For both extended and point-source-like emission, the power law exponent for brightness temperature is $\beta-2$. We note that $\alpha$ is often used as the exponent for the flux density spectral index and $\beta$ for brightness temperature. However, because in this work $\alpha$ is used for the exponent of the frequency dependence of the forward gain (Section \ref{subsec:ptsources}), we instead use $\beta$ for the flux density spectral index.

\begin{figure}[t]
\includegraphics[width=.47\textwidth]{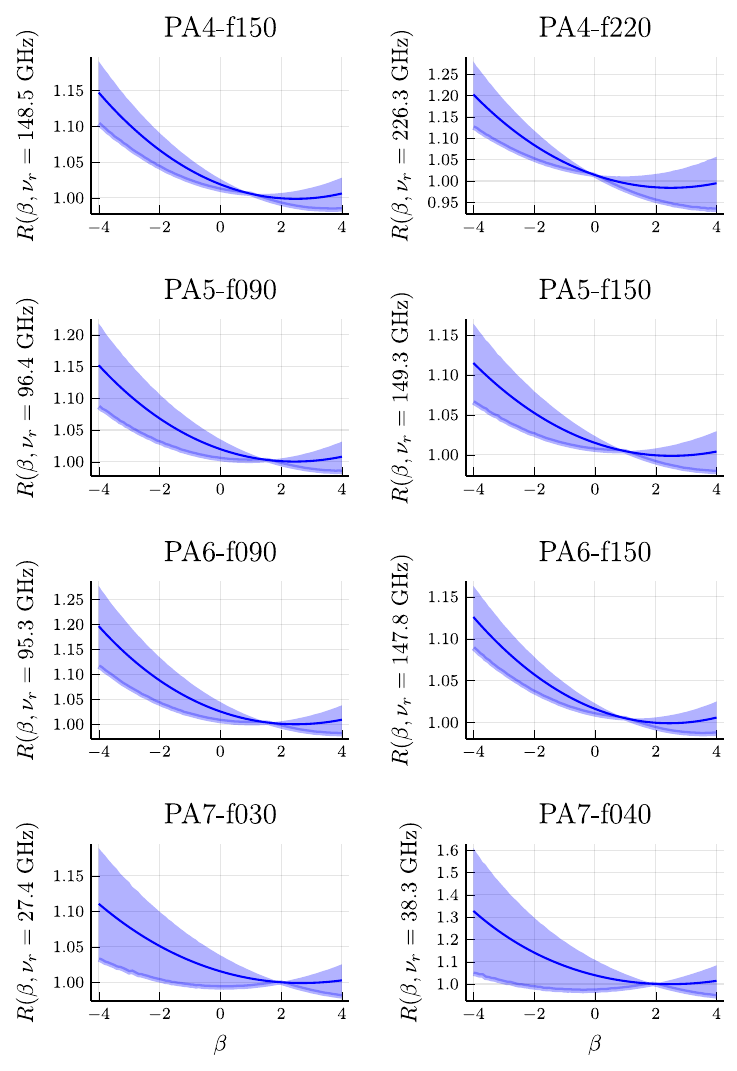}
\caption{Color correction curves of $R(\beta,\nu_r=\nu_0)$ with $1\sigma$ uncertainties for each AdvACT array. For each array, $\nu_0$ is defined by Equation \ref{eq:nu0eff} and reported in Table \ref{tab:band_centers}.}
\label{fig:color_correction_plots}
\end{figure}

\subsection{Color Corrections for Beam-Filling Sources}
\label{subsec:col_corr}

The purpose of a color correction is to convert the surface brightness of a non-CMB signal in a CMB-calibrated map to the surface brightness associated with the signal's spectral type.  
In this approach, we fix a reference frequency $\nu_r$, any one we choose, and find flux at that frequency as a function of the spectral index or equivalent. Our derivation is similar to but differs somewhat from that in \cite{planck_IX}.

The detected power is (see Equation \ref{eq:power})
\begin{equation}
P= C \int I_\nu \nu^{-2} \tau(\nu)  d\nu.
\end{equation}
Again, the $\nu^{-2}$ factor comes from the effective area for a single mode of radiation. For $I_\nu=B_\nu(T)$ in the RJ limit we recover Equation~\ref{eq:one_mode_power} with $C=c^2 / 2$. If we have a map pixel with some value $M$, then 
\begin{equation}
M=C^{\prime}\int I_\nu\nu^{-2}\tau(\nu) d\nu,
\label{eq:mdef}
\end{equation}
where $C^{\prime}$ includes the new normalization factors. 
When the map is calibrated in CMB temperature units, an input CMB signal, $\delta I_{\nu} = b_{\nu}' \delta T_{CMB}$, should yield $M = \delta T_{CMB}$. From Equation~\ref{eq:dcmb} we see that this holds true  when
\begin{equation}
C^{\prime} = \left[\int b_\nu^\prime \nu^{-2} \tau(\nu)  d\nu\right]^{-1}.
\end{equation}

Now say we have a source spectrum given by Equation~\ref{eq:dsrcdef}.
This will be measured as 
\begin{equation}
M=C^{\prime}I_{\nu_r}\int (\nu/\nu_r)^\beta\nu^{-2}\tau(\nu)  d\nu
\label{eq:cc_newsource}
\end{equation}
so that 
\begin{equation}
I_{\nu_r}= M \frac{\int b_\nu^\prime\nu^{-2} \tau(\nu) d\nu }{\int (\nu/\nu_r)^\beta \nu^{-2}  \tau(\nu) d\nu},
\label{eq:mhcc_a}
\end{equation}
which we write as
\begin{eqnarray}
M &=&\frac{I_{\nu_r}}{b_{\nu_r}^\prime} 
\frac{\int (\nu/\nu_r)^\beta\nu^{-2}  \tau(\nu)d\nu}{\int \frac{b_{\nu}^\prime}{b_{\nu_r}^\prime}\nu^{-2} \tau(\nu)d\nu }
\nonumber\\
&=& \frac{I_{\nu_r}}{b_{\nu_r}^\prime} R(\beta,\nu_r ).
\label{eq:mhcc}
\end{eqnarray}
The $R(\beta,\nu_r )$ are the ``color corrections." They convert beam-filling celestial signals, such as dust, into the CMB temperature units displayed in a map and vice versa. This last expression says pick a reasonable reference frequency $\nu_r$, take the source surface brightness $I_\nu$ at that frequency, convert it to temperature with $b_{\nu_r}^\prime$, and then multiply by $R(\beta,\nu_r )$ to put it into CMB temperature units. Plots of $R(\beta,\nu_r )$ for each of the AdvACT arrays are shown in Figure \ref{fig:color_correction_plots}. The $1\sigma$ uncertainties are obtained using the same Monte Carlo process detailed in \ref{subsubsec:error_prop}.

\subsection{Effective Frequencies for Beam-Filling Sources}

\paragraph{Central frequency}
The independence of frequency in Equation \ref{eq:one_mode_power} is a consequence of being in the RJ regime, but for any instrument we have a general idea of the frequency of observation. It is somewhere near the middle of the band. However, there is no physical principle that determines its value. Therefore, we adopt the following definition for the central frequency for a receiver viewing a beam-filling RJ source, the same as used in the main body of this work, Equation \ref{eq:RJ_center_freq}:
\begin{equation}
\nu_0 \equiv \frac{\int \nu \tau(\nu ) d\nu}{\int \tau(\nu )d\nu}.
\label{eq:nu0eff}
\end{equation}
For flat passband between $\nu_a$ and $\nu_b$, $\nu_0=(\nu_a+\nu_b)/2$
in accord with our intuition.

\paragraph{Effective frequencies}
Jarosik et al. \cite{Jarosik_2003} give the frequency at which a beam-filling source should be evaluated so that its emission equals the band averaged quantity. We now derive this effective frequency for a source characterized by $I_\nu$.
In a small band $\delta\nu$,
\begin{equation}
P = I_\nu \frac{c ^2}{\nu ^2}\tau(\nu )\delta\nu \rightarrow 
\frac{\int I_\nu \frac{c^2}{\nu^2}\tau(\nu )d\nu}{\int \tau(\nu) d\nu}\Delta\nu.
\label{eq:bbnu1}
\end{equation}

The expression to the right of the arrow simply sums up all the sub-band contributions to make the full band, $\Delta\nu$. This expression is the same as Equation~49 in \cite{Jarosik_2003}, up to some constants and a different convention for $\beta$. The power detected in one spatial mode in both polarizations can be expressed in terms of the brightness temperature as
\begin{equation}
P = 2k_BT\Delta\nu = 2k_B\frac{I_{\nu_{e}}t_{\nu_e}^\prime}{b_{\nu_e}^\prime}\Delta\nu,
\label{eq:bbnu2}
\end{equation}
where $\nu_e$ is the effective frequency of the source.
Note that  $b_\nu^\prime/t_\nu^\prime = b_0x^2$. In the RJ limit $I\rightarrow b_0Tx^2= 2\nu^2k_BT/c^2$ and Equations~\ref{eq:bbnu1} and \ref{eq:bbnu2} give the same thing. As is the case in \cite{Jarosik_2003}, in the RJ limit for a beam-filling source frequency drops out of both sides and so the effective frequency is set by Equation~\ref{eq:nu0eff}. For other sources we solve the following equation to determine the effective frequency
\begin{equation}
\frac{I_{\nu_{e}}t_{\nu_e}^\prime}{b_{\nu_e}^\prime} \equiv \frac{c^2}{2k_B} \frac{\int I_\nu \nu^{-2}\tau(\nu )d\nu}{\int \tau(\nu) d\nu}.
\label{eq:bbnueff}
\end{equation}

For a map calibrated in $\delta T_{CMB}$ we may plug in Equation \ref{eq:dcmb} to find that the effective frequency is defined by
\begin{equation}
2k_B \delta T_{CMB}t_{\nu_e}^\prime
=\delta T_{CMB} \frac{\int b_\nu^\prime \frac{c^2}{\nu^2}\tau(\nu )d\nu}{\int \tau(\nu) d\nu},
\label{eq:ttnueff}
\end{equation}
or 
\begin{equation}
t_{\nu_c}^\prime =  \frac{c^2}{2k_B } \frac{\int b_\nu^\prime  \nu^{-2}\tau(\nu )d\nu}{\int \tau(\nu) d\nu},
\label{eq:nuceff}
\end{equation}
where we have let $\nu_e=\nu_{c}$ to denote the CMB. The values for $\nu_c$ and for $\nu_e$ for a variety of sources are given in Appendix~\ref{appen:results}.

\paragraph{Effective frequencies vs. color corrections for beam-filling sources.}

Suppose that we have a map calibrated in $\delta T_{CMB}$ at $\nu_c$ and we measure a source, such as the galactic plane, with a spectrum given by
Equation~\ref{eq:dsrcdef}. What do we report for $I_{\nu_r}$? With $I_\nu$ (Equation~\ref{eq:dcmb}) evaluated at the CMB effective frequency, $\nu_c$, and Equation~\ref{eq:bbnueff} we get
\begin{equation}
I_{\nu_{c}} =  \delta T_{CMB}b_{\nu_{c}}^\prime =  \frac{c^2}{2k_B}  \frac{b_{\nu_{c}}^\prime}{t_{\nu_{c}}^\prime} \frac{\int I_{\nu_r} \left(\frac{\nu}{\nu_r}\right)^\beta \nu^{-2}\tau(\nu )d\nu}{\int \tau(\nu) d\nu}.
\label{eq:ttnueff3}
\end{equation}

Solving for $I_{\nu_r}$ and substituting in Equation~\ref{eq:nuceff} for $t_{\nu_c}^\prime$, we find 
\begin{equation}
I_{\nu_r} = I_{\nu_{c}} \frac{1}{b_{\nu_{c}}^\prime} \frac{\int b_{\nu}^\prime \nu^{-2}\tau(\nu) d\nu}{\int \left(\frac{\nu}{\nu_r}\right)^\beta \nu^{-2}\tau(\nu )d\nu}.
\end{equation}

Because $M=I_{\nu_{c}}/b_{\nu_{c}}^\prime$ the above is the same as Equation~\ref{eq:mhcc_a}. Thus, the color correction and effective frequency approaches encode the same information but in different ways. The color corrections, though, are often more convenient because they give the response for different values of $\beta$ at a reference frequency as opposed to a series of effective frequencies for different values of $\beta$. We note that in the ACT DR6 power spectrum analysis, neither approach is adopted explicitly. Instead, the frequency response of each component is integrated directly over the measured passbands at each theory-model evaluation.

\subsection{Effective Frequencies for Point Sources}
\label{subsec:ptsources}
Point sources appear in our maps as hot spots with the profile of the instrument beam.\footnote{In this treatment we ignore the pixel window function.} Their amplitudes are calibrated to the CMB, while the beam profile is measured using planet scans. The source flux, $S_\nu$, is related to the surface brightness through
\begin{equation}
I_\nu = S_\nu\delta(\cos\theta -\cos \theta_0)\delta(\phi-\phi_0) ,
\end{equation}
where the delta functions pick out its position at coordinates $(\theta_0,\phi_0)$. The power measured in small band $\delta\nu$ while pointing in direction $\vec\theta=(\theta,\phi)$ is
\begin{equation}
P(\vec\theta) = \int I_\nu(\vec\theta^\prime-\vec\theta) A_{e}(\hat{n^\prime},\nu)\,d\Omega_\nu^\prime~\delta\nu,
\end{equation}
where, as before, $A_{e}$ is the effective area of the primary optic. The integral is over the solid angle. For ease of notation we have suppressed a possible rotation of the beam around its axis. By the reciprocity theorem or energy conservation (see e.g., \cite{kraus_radio_astronomy_1986} or \cite{condon_ransom_2016}), $A_{e}(\hat{n},\nu)=\lambda^2G_\nu(\theta,\phi)/4\pi$, the more general form of ``$A\Omega=m\lambda^2$" with $m=1$ for a single mode. With this, 
\begin{equation}
P(\vec\theta) = \int I_\nu(\vec\theta^\prime-\vec\theta) \frac{c^2}{4\pi \nu^2}G_\nu( \vec\theta^\prime) d\Omega_\nu^\prime~\delta\nu.
\end{equation}
Combining the contributions from each sub-band weighted by the bandwidth, we obtain
\begin{equation}
P =  2k_BT\Delta\nu= \frac{c^2}{4\pi}\frac{ \iint I_\nu(\vec\theta^\prime-\vec\theta) \nu^{-2}  G_\nu(\vec\theta^\prime)  \tau(\nu) d\Omega_\nu^\prime d\nu}{\int \tau(\nu) d\nu}\Delta\nu.
\end{equation}
Integrating over solid angle and substituting in for $I_\nu$ we find \cite{Page_2003}
\begin{equation}
T_r = \frac{c^2S_{\nu_r}}{8\pi k_B\nu_r^\beta} \frac{ \int \nu^{\beta}\nu^{-2}   G_m(\nu)\tau(\nu) d\nu}{\int \tau(\nu) d\nu}=\Gamma S_{\nu_r},
\label{eq:tsource}
\end{equation}
where $G_m(\nu)$ is the maximum forward gain.
It is defined through 
\begin{equation}
G_\nu(\theta,\phi) = G_m(\nu ) B(\theta, \phi)= 
\frac {4\pi }{\Omega_B}\bigg(\frac{\nu}{\nu_{B}}\bigg)^{\alpha}B(\theta, \phi),
\end{equation}
where $B$ is the normalized beam profile and $\nu_B$ is the frequency at which the beam solid angle, $\Omega_B$, is measured. Here we have modeled the frequency dependence of the forward gain as $G(\nu) = G_{\nu_B}(\nu/\nu_B)^{\alpha}$. When $\alpha = 0$, the beam is independent of frequency.  With regard to what a detector measures, this is the same as one gets with a beam-filling source. When $\alpha = 2$, the illuminated area of the primary is independent of frequency. This behavior is expected because the cryogenic Lyot stop is imaged to a fixed radius on the primary \cite{Fowler_2007}. 

Combining the previous two equations gives
\begin{equation}
\Gamma = \frac{c^2}{2 k_B\Omega_B\nu_r^\beta \nu_{B}^{\alpha}}
\frac{\int \nu^{\beta}\nu^{\alpha-2}\tau(\nu)\,d\nu}
{\int \tau(\nu)\,d\nu}
\,10^{-20}
\quad [\mu\mathrm{K}/\mathrm{Jy}],
\label{eq:gamma}
\end{equation}
for any reference frequency $\nu_r$.
The $\nu^{-2}$ in the numerator is grouped with the $\alpha$ because both are associated with the gain. By convention, $\Gamma$ converts the power in both polarizations to a brightness temperature.

To find an effective frequency for a source, we let $T_e=T_r$. These can be written as 

\begin{eqnarray}
T_r &=& \frac{c^2 S_{\nu_r}}{2 k_B\Omega_B\nu_r^\beta \nu_{B}^{\alpha} } \frac{ \int \nu^{\beta} \nu^{\alpha-2} \tau(\nu)  d\nu}{\int \tau(\nu) d\nu},\\
T_e &=& \frac{c^2 S_{\nu_r}}{2 k_B\Omega_B} \left(\frac{\nu_e}{\nu_r}\right)^{\beta} \left(\frac{\nu_e}{\nu_B}\right)^{\alpha} {\nu_e}^{-2}.
\end{eqnarray}

Solving for $\nu_e$ we find \footnote{This updates equation 26 in Page et al 2003 \cite{Page_2003} The correction was pointed out to L. Page by Ricardo Genova.}

\begin{equation}
\nu_e^{(\beta+\alpha-2)} =  \frac{ \int \nu^{\beta}\nu^{\alpha-2}  \tau(\nu) d\nu}{\int \tau(\nu) d\nu}.
\label{eq:nueff_ps}
\end{equation} 
For a beam-filling ($\alpha=0$) source, this reduces to the result from Equation \ref{eq:bbnueff}.


From Equation \ref{eq:nueff_ps}, we see that given the effective frequency and solid angle, we can compute the flux at any reference frequency with
\begin{equation}
\Gamma = \frac{c^2}{2k_B\Omega_B\nu_e^2}\left( \frac{\nu_e}{\nu_B}  \right)^{\alpha} \left( \frac{\nu_e}{\nu_r}  \right)^{\beta}.
\end{equation}
For $\nu_r=\nu_e$, this reduces to

\begin{equation}
\Gamma = \frac{c^2}{2k_B\Omega_B\nu_e^2} \left( \frac{\nu_e}{\nu_B}  \right)^{\alpha} = \frac{c^2}{2k_B \nu_e^2}\frac{G_m(\nu_e)}{4 \pi} = \frac{A_e}{2 k_B},
\end{equation} 
which is equation 3.48 of \cite{condon_ransom_2016}. Here the effective area is computed at the effective frequency. 
Appendix~\ref{appen:results} gives $\nu_e$, $\Gamma$, and $\Omega_B$ in a variety of cases. A convenient form for $\Gamma$ at a fixed frequency is $\Gamma = (b_0x^2\Omega_B)^{-1}$.

\paragraph{Connecting ACT's measurements to the effective frequency.}

Point sources are detected in our maps with a matched filter that returns the amplitude of the source in units of $\delta T_{CMB}$. The source has some effective frequency, $\nu_e$, in the band and shows up in the map with a brightness temperature
\begin{equation}
T_{s,\nu_e}  = \delta T_{s, CMB}t_{\nu_{e}}^\prime
\end{equation}
and an associated flux $S_{\nu_e}  = T_{s,\nu_e}/\Gamma_{\nu_e}$. In general, $\nu_e$ is different than $\nu_c$.
From here we can compute the flux at any other reference frequency with
$S_{\nu_r}  = S_{\nu_e}\left(\frac{\nu_r}{\nu_e}\right)^{\beta}$.

The prescription for reporting the flux from a point source with known $\beta$ at, for example, $\nu_r=150~$GHz, is then a) use a matched filter to find $\delta T_{s, CMB}$, b) compute the effective frequency for $\nu_r$, $\beta$, $\alpha$, and $\tau(\nu)$
from Equation~\ref{eq:nueff_ps}, c) compute $t_{\nu_e}^\prime$ from Equation~\ref{eq:tprime} to get the brightness temperature $T_{s,\nu_e}$ from $\delta T_{s, CMB}$, d) compute $\Gamma$ for $\nu_e$ and $\nu_r$ to find the flux, $S_{\nu_e}$, from $T_{s,\nu_e}$, and e) find the flux at the reference frequency, $S_{\nu_r}$.



\onecolumn
\section{Numerical Results}
\label{appen:results}

We present the general characteristics of all the passbands measured for ACT in the common framework outlined in this work. The quoted uncertainties are based on the analysis described in this paper and do not include potential systematic shifts from out of band features at or below our noise level.

\begin{table*}[ht]
\caption{\centering{Passband characteristics for MBAC}}
\renewcommand{\arraystretch}{0.8} 
\vspace{-0.1in}
\begin{center}
\begin{tabular}{l c c c  }
\hline
\hline
Array  & AR1 (GHz) & AR2 (GHz)& AR3 (GHz) \\
 \hline
$\nu_0$   & $150.1\pm3.5$ & $220.4\pm3.5$ & $276.5\pm3.5$ \\
Bandwidth, $\Delta\nu$   & 18.2 & 17.0 & 20.5\\
Dicke bandwidth, $\Delta_D\nu$   & 28.3 & 25.8 & 35.4 \\
\hline
\multicolumn{4}{c}{Effective frequencies for beam-filling sources}\\
\hline
Synchrotron ($\beta=-0.7$)  &149.1 &219.8  & 276.0 \\
Free-free ($\beta=-0.1$)  & 149.3 & 219.9  & 276.0 \\
Dust ($\beta=3.5$)  & 150.2 & 220.5 &276.7 \\
CMB ($\nu_c$) &150.0 & 220.2  & 276.0 \\
\hline
\multicolumn{4}{c}{$\Gamma$ and effective frequencies for point sources}\\
\hline
$\Gamma$ ($\mu$K/Jy), ($\nu_{e}$)  ($\beta=-0.7$)  & 6460 (149.6) & 5400 (220.0)  & 4340 (276.2) \\
$\Gamma$ ($\mu$K/Jy), ($\nu_{e}$)  ($\beta=-0.1$)  & 6460 (149.7) & 5390 (220.1)   &  4340 (276.3)\\
$\Gamma$ ($\mu$K/Jy), ($\nu_{e}$)  ($\beta=3.5$)  & 6450 (150.6) & 5380 (220.6) & 4330 (277.2) \\

\hline
\multicolumn{4}{c}{Conversion factors and beam characteristics}\\
\hline
$\delta T_{CMB}/\delta T_{RJ}$ at $\nu=\nu_c$ & $1.74\pm0.04$ & $3.08\pm0.10$ & $5.38\pm0.20$ \\
$\delta W/\delta T_{RJ}$ (pW/K) & 0.503 & 0.469 & 0.566 \\
$\Omega_B$ (nsr) & $223.2\pm 4.0$ & $124.4\pm2.5$ & $98.1\pm2.0$ \\
$\nu_B$ (GHz) & 150.3 & 220.4 & 276.5 \\
$\alpha$  & 1.8 & 1.1 & 1.1 \\ 
\hline
\hline
\end{tabular}
\end{center}
\vspace{-0.1in} 
\small{This table updates Table 4 in \cite{Swetz_2011}. The central values have shifted somewhat within the original uncertainty but we retain the original value of $\pm3.5~$GHz on frequencies and retain other error bars as well.
The $\delta T_{RJ}$ to $\delta T_{CMB}$ conversion is the inverse of Equation~\ref{eq:tprime}.
We now report $\delta W/\delta T_{RJ}$ for $\eta=1$ in 
Equation~\ref{eq:one_mode_power} but for two polarizations because we measure both. The uncertainties on $\Gamma$ are 320, 200 and 140 $\mu$K/Jy for AR1, AR2, and AR3 respectively. For $\Omega_B$ we use the weighted average of the values in Table 2 of \cite{Hasselfield_2013} increased to account for beam jitter reported in Table 3 therein. We have also increased the uncertainties to account for systematic errors described in \cite{Lungu_2022}. The $\nu_B$ is from Equation~\ref{eq:nueff_ps} using estimates of $\alpha$ based on subsequent modeling of the ACT forward gain via GRASP \footnotemark.} 
\label{tab:mbac_freqs}
\end{table*}

\footnotetext{GRASP. 2018, The GRASP reflector antenna design software package is available through \url{https://www.ticra.com}.}

\begin{table*}[ht]
\caption{\centering{Passband characteristics for ACTPol}}
\renewcommand{\arraystretch}{0.80} 
\vspace{-0.1in}
\begin{center}
\begin{tabular}{l c c c c }
\hline
\hline
Array  & PA1 (GHz) & PA2 (GHz)& PA3, f090 (GHz) & PA3, f150 (GHz) \\
 \hline
$\nu_0$   & $150.9\pm2.4$ & $151.3\pm2.4$ & $98.4\pm1.7$ & $148.5\pm2.4$\\
Bandwidth, $\Delta\nu$   & $33.5\pm2.1$ & $33.0\pm1.9$ & $20.7\pm1.9$ & $30.1\pm1.6$\\
Dicke bandwidth, $\Delta_D\nu$   & $50.3\pm1.9$ & $51.0\pm1.9$ & $34.1\pm1.9$ & $42.2\pm1.6$\\
\hline
\multicolumn{5}{c}{Effective frequencies for beam-filling sources}\\
\hline
Synchrotron ($\beta=-0.7$)  &148.4 &148.8  & 96.2 & 146.8 \\
Free-free ($\beta=-0.1$)  & 148.8 & 149.2  & 96.6 &  147.1\\
Dust ($\beta=3.5$)  & 151.2 & 151.6 & 98.7 & 148.7\\
CMB ($\nu_c$) &150.8 & 151.2  & 98.7 & 148.4\\
\hline
\multicolumn{5}{c}{$\Gamma$ and effective frequencies for point sources}\\
\hline
$\Gamma$ ($\mu$K/Jy), ($\nu_{e}$)  ($\beta=-0.7$)  &7080 (149.7) & 7460 (150.0)  & 6570 (97.4)  & 5550 (147.7) \\
$\Gamma$ ($\mu$K/Jy), ($\nu_{e}$)  ($\beta=-0.1$)  & 7080 (150.1) & 7450 (150.4) & 6570 (97.7)  &  5550 (147.9)\\
$\Gamma$ ($\mu$K/Jy), ($\nu_{e}$)  ($\beta=3.5$)  & 7060 (152.4) & 7430 (152.8) & 6560 (99.7)  & 5540 (149.5)\\

\hline
\multicolumn{5}{c}{Conversion factors and beam characteristics}\\
\hline
$\delta T_{CMB}/\delta T_{RJ}$ at $\nu=\nu_c$ & $1.75\pm0.03$ & $1.75\pm0.03$ &  $1.28\pm0.01$ & $1.72\pm0.03$\\
$\delta W/\delta T_{RJ}$ (pW/K) & 0.463 & 0.456 & 0.286 & 0.416 \\
$\Omega_B$ (nsr) & $200.8\pm5.0$ & $189.9\pm4.3$ & $507\pm22$ & $265.0\pm6.4$\\
$\nu_B$   & $151.4$ & $151.8$ & $98.9$ & $148.9$\\
$\alpha$  & 1.8 & 1.8 & 1.9 & 1.8\\ 
\hline
\hline
\end{tabular}
\end{center}
\vspace{-0.1in} 
\small{This updates Table~1 in \cite{Thornton_2016} and Tables A1 and A2 in \cite{Datta_2019}. The central values have shifted by roughly the original quoted uncertainties (2.4 GHz) and we retain the original error bound of $\pm 2.4$ GHz on f150 bands and $\pm$ 1.7 GHz on the f090 band. The uncertainties on the $\Gamma$'s are 290, 290, 360, and 230 $\mu$K/Jy for PA1, PA2, PA3 f090, and PA3 f150, respectively.
The $\delta T_{RJ}$ to $\delta T_{CMB}$ conversion is the inverse of Equation~\ref{eq:tprime}. We give $\delta W/\delta T_{RJ}$ for $\eta=1$ in 
Equation~\ref{eq:one_mode_power} for one polarization.
The values for $\alpha$ are estimated based on modeling of the ACT forward gain using GRASP. For $\Omega_B$, we use the weighted average over all observations in Table 4 of \cite{Lungu_2022}. These solid angles include the ``jitter beam.'' The $\nu_B$ is from Equation~\ref{eq:nueff_ps}.}
\label{tab:apol_freqs}
\end{table*}

\begin{table*}[ht]
\caption{\centering{Passband characteristics for Advanced ACTPol, part A}}
\renewcommand{\arraystretch}{1.0} 
\vspace{-0.1in}
\begin{center}
\begin{tabular}{l c c c c c}
\hline
\hline
Array  & PA7, f030 (GHz) & PA7, f040 (GHz)& PA5, f090 (GHz) & PA6, f090 (GHz) & 
PA4, f220 (GHz) 
\\
 \hline
$\nu_0$  & $27.4\pm0.3$ & $38.3\pm1.3$ & $96.4\pm0.9$ & $95.3\pm1.1$ & $226.1\pm3.5$ \\
Bandwidth, $\Delta\nu$   & $4.3\pm0.4$ & $12.4\pm0.8$  & $18.9\pm1.0$ & $23.1\pm1.0$ & $64.9\pm2.5$\\
Dicke bandwidth, $\Delta_D\nu$ & $6.3\pm0.2$ & $15.6\pm0.3$ & $26.9\pm0.3$ & $30.3\pm0.3$ & $81.4\pm0.6$\\
\hline
\multicolumn{6}{c}{Effective frequencies for beam-filling sources}\\
\hline
Synchrotron ($\beta=-0.7$)  &27.1&37.4 &95.3  & 93.9 & 222.1 \\
Free-free ($\beta=-0.1$)  &27.2& 37.6 & 95.5  & 94.1 &  222.7\\
Dust ($\beta=3.5$)  &27.4& 38.4 & 96.6 & 95.5 & 226.6\\
CMB ($\nu_c$) &27.4 & 38.5& 96.6  & 95.5 & 224.5\\
\hline

\multicolumn{6}{c}{$\Gamma$ and effective frequencies for point sources}\\
\hline
$\Gamma$ ($\mu$K/Jy), ($\nu_{e}$) ($\beta=-0.7$)  & 8570 (27.3)& 7850 (37.9) & 7170 (95.9)  & 7290 (94.6) & 5800 (223.3) \\
$\Gamma$ ($\mu$K/Jy), ($\nu_{e}$)  ($\beta=-0.1$)  & 8570 (27.3) & 7850 (38.0)  & 7170 (96.0)  & 7290 (94.9) &  5790 (223.9)\\
$\Gamma$ ($\mu$K/Jy), ($\nu_{e}$) ($\beta=3.5$)  & 8570 (27.5)& 7840 (38.9) & 7160 (97.2) & 7280 (96.2) 
& 5700 (227.8)\\

\hline
\multicolumn{6}{c}{Conversion factors and beam characteristics}\\
\hline
$\delta T_{CMB}/\delta T_{RJ}$ at $\nu=\nu_c$ &$1.02 \pm 0.01$& $1.04\pm 0.01$ & $1.27 \pm 0.01$ & $1.26 \pm 0.01$& $3.21\pm 0.10$\\
$\delta W/\delta T_{RJ}$ (pW/K) & 0.059 & 0.171 & 0.261 & 0.319 & 0.900\\
$\Omega_B$ (nsr) & $5060^{a}$ & $2800^{a}$  & $487.8\pm3.6$ & $491.9\pm2.6$ & $110.6\pm1.6$\\
$\nu_B$   & 27.4 & 38.5 & 96.5 & 95.3 & 226.7\\
$\alpha$  & 1.99 & 1.94 & 1.9 & 1.9 & 1.1\\ 
\hline
\hline
\end{tabular}
\end{center}

\vspace{-0.1in} 
\small{The uncertainties on the $\Gamma$'s are 250, 550, 150, 180, and 180 $\mu$K/Jy for arrays PA7 f030, PA7 f040, PA5 f090, PA6 f090, and PA4 f220, respectively.
The $\delta T_{RJ}$ to $\delta T_{CMB}$ conversion is the inverse of Equation~\ref{eq:tprime}. We give $\delta W/\delta T_{RJ}$ for $\eta=1$ in 
Equation~\ref{eq:one_mode_power} for one polarization.
For $\Omega_B$, we use the solid angles computed from estimates of the monochromatic beam profile evaluated at each reference frequency $\nu_B$. These solid angles include the ``jitter beam.'' The profiles were released through NASA’s Legacy Archive for Microwave Background Data Analysis (LAMBDA; \url{https://lambda.gsfc.nasa.gov/product/act/act_dr6.02/act_dr6.02_harmonic_beams_profiles_info.html}) and are described as part of the ACT DR6 map release \cite{Naess_2025}. The files follow the naming convention \path{act_dr6.02_main_beams/main_beams/nominal/coadd_{array}_{band}_night_beam_tform_jitter_mono.txt}, where \path{{array}} and \path{{band}} denote the detector array and frequency band listed in the corresponding table column (e.g., \path{pa5} and \path{f090}). The solid angles for PA7 are preliminary and not a data product. They are included to indicate the expectations from PA7.}
\label{tab:aaact_pta_freqs}
\end{table*}

\begin{table*}[ht]
\caption{\centering{Passband characteristics for Advanced ACTPol, part B}}
\vspace{-0.1in}
\renewcommand{\arraystretch}{1.0} 
\begin{center}
\begin{tabular}{l c c c }
\hline
\hline
Array  & PA4, f150 (GHz) & PA5, f150 (GHz)& PA6, f150 (GHz) \\
 \hline
$\nu_0$  & $148.5\pm1.1$ & $149.3\pm1.3$ & $147.8\pm0.9$  \\
Bandwidth, $\Delta\nu$   & $35.8\pm1.4$ & $28.1\pm1.4$  & $31.2\pm1.4$\\
Dicke bandwidth, $\Delta_D\nu$ & $44.2\pm0.3$ & $38.2\pm0.3$ & $40.5\pm0.3$ \\
\hline
\multicolumn{4}{c}{Effective frequencies for beam-filling sources}\\
\hline
Synchrotron ($\beta=-0.7$)  &146.7 &147.9 &146.3   \\
Free-free ($\beta=-0.1$)  &147.0& 148.1 & 146.5  \\
Dust ($\beta=3.5$)  &148.8& 149.5 & 148.0 \\
CMB ($\nu_c$) &148.4 & 149.2& 147.8 \\
\hline
\multicolumn{4}{c}{$\Gamma$ and effective frequencies for point sources}\\
\hline
$\Gamma$ ($\mu$K/Jy), ($\nu_{e}$) ($\beta=-0.7$)  & 6470 (147.6) & 6770 (148.6) & 6710 (147.0)   \\
$\Gamma$ ($\mu$K/Jy), ($\nu_{e}$) ($\beta=-0.1$)  & 6470 (147.9)&  6770 (148.8) & 6710 (147.3)  \\
$\Gamma$ ($\mu$K/Jy), ($\nu_{e}$) ($\beta=3.5$)  & 6460 (149.6)& 6760 (150.1) &  6700 (148.8) \\
\hline
\multicolumn{4}{c}{Conversion factors and beam characteristics}\\
\hline
$\delta T_{CMB}/\delta T_{RJ}$ at $\nu=\nu_c$ &$1.72\pm 0.01$ & $1.73 \pm 0.02$ & $1.71 \pm 0.01$\\
$\delta W/\delta T_{RJ}$ (pW/K)  & 0.494 & 0.388 & 0.429 \\
$\Omega_B$ (nsr) & $228.3\pm1.9$ & $215.8\pm1.5$ & $221.9\pm1.1$\\
$\nu_B$   & 148.5 & 149.3 & 147.9 \\
$\alpha$  & 1.8 & 1.8 & 1.8 \\ 
\hline
\hline
\end{tabular}
\end{center}
\vspace{-0.1in} 
\small{The uncertainties on the $\Gamma$'s are 100, 120, and 90 $\mu$K/Jy for arrays PA4, PA5, and PA6, respectively.
The $\delta T_{RJ}$ to $\delta T_{CMB}$ conversion is the inverse of Equation~\ref{eq:tprime}. We give $\delta W/\delta T_{RJ}$ for $\eta=1$ in 
Equation~\ref{eq:one_mode_power} for one polarization.
For $\Omega_B$, we use the solid angles computed from estimates of the monochromatic beam profile evaluated at each reference frequency $\nu_B$. These solid angles include the ``jitter beam.'' The profiles were released through NASA’s Legacy Archive for Microwave Background Data Analysis (LAMBDA; \url{https://lambda.gsfc.nasa.gov/product/act/act_dr6.02/act_dr6.02_harmonic_beams_profiles_info.html}) and are described as part of the ACT DR6 map release \cite{Naess_2025}. The files follow the naming convention \path{act_dr6.02_main_beams/main_beams/nominal/coadd_{array}_{band}_night_beam_tform_jitter_mono.txt}, where \path{{array}} and \path{{band}} denote the detector array and frequency band listed in the corresponding table column (e.g., \path{pa5} and \path{f150}).}
\label{tab:aaact_ptb_freqs}
\end{table*}

\FloatBarrier

\twocolumn

\begin{backmatter}

\bmsection{Acknowledgment}
Support for ACT was through the U.S.~National Science Foundation through awards AST-0408698, AST-0965625, and AST-1440226 for the ACT project, as well as awards PHY-0355328, PHY-0855887 and PHY-1214379. Funding was also provided by Princeton University, the University of Pennsylvania, and a Canada Foundation for Innovation (CFI) award to UBC. ACT operated in the Parque Astron\'omico Atacama in northern Chile under the auspices of the Agencia Nacional de Investigaci\'on y Desarrollo (ANID). The development of multichroic detectors and lenses was supported by NASA grants NNX13AE56G and NNX14AB58G. Detector research at NIST was supported by the NIST Innovations in Measurement Science program. Computing for ACT was performed using the Princeton Research Computing resources at Princeton University, the National Energy Research Scientific Computing Center (NERSC), and the Niagara supercomputer at the SciNet HPC Consortium. SciNet is funded by the CFI under the auspices of Compute Canada, the Government of Ontario, the Ontario Research Fund–Research Excellence, and the University of Toronto. We thank the Republic of Chile for hosting ACT in the northern Atacama, and the local indigenous Licanantay communities whom we follow in observing and learning from the night sky. R.D. thanks ANID for grants BASAL CATA FB210003, QUIMAL-240004 and FONDECYT-1262583. E.C. acknowledges support from the European Research Council (ERC) under the European Union’s Horizon 2020 research and innovation programme (Grant agreement No. 849169).

\bmsection{Disclosures}
The authors declare no conflicts of interest.

\bmsection{Data availability}
Data underlying the simulations presented in this work are available on Zenodo \cite{alford_2026_FTS}. The codes used to generate the FTS optical simulations (FTS-Simulation; \cite{Alford_FTS_Simulation}) and process the measured passbands (passband-analysis; \cite{Alford_Passband_Analysis}) are available on GitHub.

\end{backmatter}

\bibliography{sample}

\bibliographyfullrefs{sample}

\end{document}